\documentclass[prb,aps,twocolumn,superscriptaddress,showpacs]{revtex4-1}
\usepackage[colorlinks=true,linkcolor=black, citecolor=blue, urlcolor=blue, 
    unicode=true]{hyperref}

\usepackage{graphicx}
\usepackage{amsmath,amssymb} 
\usepackage{physics}

\begin{document}

\title{Imitation of spin density wave order in Cu$_3$Nb$_2$O$_8$}

\author{N. Giles-Donovan}
\affiliation{Centre for Medical and Industrial Ultrasonics, School of Engineering, University of Glasgow G12 8QQ, UK}

\author{N. Qureshi}
\affiliation{Institut Laue-Langevin, 71 avenue des Martyrs, CS 20156, 38042 Grenoble Cedex 9, France}

\author{R. D. Johnson}
\affiliation{Department of Physics and Astronomy, University College London, Gower Street, London, WC1E 6BT}

\author{L. Y. Zhang}
\affiliation{Laboratory for Pohang Emergent Materials, Pohang Accelerator Laboratory and Max Plank POSTECH Center for Complex Phase Materials, Pohang University of Science and Technology, Pohang 790-784, Korea}

\author{S.-W. Cheong}
\affiliation{Rutgers Center for Emergent Materials and Department of Physics and Astronomy, Rutgers University, 136 Frelinghuysen Road, Piscataway, New Jersey 08854, USA}

\author{S. Cochran}
\affiliation{Centre for Medical and Industrial Ultrasonics, School of Engineering, University of Glasgow G12 8QQ, UK}

\author{C. Stock}
\affiliation{School of Physics and Astronomy, University of Edinburgh, Edinburgh EH9 3JZ, UK}

\date{\today}

\begin{abstract}
Spin density waves, based on modulated local moments, are usually associated with metallic materials, but have recently been reported in insulators which display coupled magnetic and structural order parameters. We discuss one such example, the multiferroic Cu$_3$Nb$_2$O$_8$, which is reported to undergo two magnetic phase transitions, first to a spin density wave phase at $T_N \approx 26.5K$, and then to a helicoidal  structure coupled to an electric polarization below $T_2 \approx 24K$ [R. D. Johnson, \textit{et al.}, Phys. Rev. Lett., {\bf 107}, 137205 (2011)] which breaks the crystallographic inversion symmetry. We apply spherical polarimetry to confirm the low-temperature magnetic structure, yet only observe a single magnetic phase transition to helicoidal order.  We argue that the reported spin density wave originates from a decoupling of the components of the magnetic order parameter, as allowed by symmetry and driven by thermal fluctuations.  This provides a mechanism for the magnetic, but not nuclear, structure to break inversion symmetry thereby creating an intermediate phase where the structure imitates a spin density wave. As the temperature is reduced, this intermediate structure destabilizes the crystal such that a structural chirality is induced, as reflected by the emergence of the electric polarization, and the imitation spin density wave relaxes into a generic helicoid.  This provides a situation where the magnetic structure breaks inversion symmetry while the crystal structure remains centrosymmetric.
\end{abstract}


\maketitle

\section{Introduction}

A multiferroic displays a coupling between two `ferroic' orders in the same phase and this has usually been limited to intertwined magnetic and electrical order.\cite{Spaldin_Review,Cheong_Review,Spaldin_Ramesh_Review} The ability to couple ferromagnetic and ferroelectric order parameters at workable temperatures would open up the ability to create new magnetoelectric devices independently controllable with magnetic and electric fields.\cite{Kimura_TMO} However, magnetism favors empty orbitals while ferroelectricity tends to require filled orbitals therefore making such coupling difficult to realize in materials. This requires the exploration of new mechanisms utilizing, for example, crystalline symmetry. Due to the delicate competition between different order parameters, multiferroics provide the opportunity for potentially novel structural and magnetic ground states with relaxors~\cite{Ye98:155,Cowley11:60,Stock20:xx} and skyrmion phases~\cite{Fert17:2} being just several of many examples.  

Multiferroics, where one of the order parameters of interest is electric polarization, are typically insulators~\cite{Anderson65:14,Shi13:12,Neill17:95}, however, recently non metallic systems have been reported where spin density wave like structures have been found~\cite{NVO_1} to coexist with ferroelectricity. In a metallic material, a spin density wave is defined as a modulation of the local moment and arises due the presence of a nesting vector which links parts of the Fermi surface.\cite{SDWRef1,SDWRef2} However, insulators lack such a surface and so these spin density waves must be the product of a different mechanism, or, be indicative of an alternative structure which only emulates a spin density wave.

Cu$_3$Nb$_2$O$_8$ (room temperature symmetry $P\bar{1}$) experiences two phase transitions at low temperatures: it magnetically orders at $T_N \approx 26.5K$ with incommensurate propagation vector $\vec{k} = (0.4876, 0.2813, 0.2029)$ (referred to here as the middle temperature - MT - phase) and develops an electric polarization along the real space direction $[1, 3, 2]$ below $T_2 \approx 24K$ (low temperature - LT - phase). \cite{Johnson_CNO_Paper} Johnson \textit{et al.} reported that the LT phase has a chiral structure allowed by the breaking of the single inversion center ($P\bar{1} \rightarrow P1$) during the transition.     

\begin{figure}[h!]
	\centering
	\includegraphics[width=.8\linewidth]{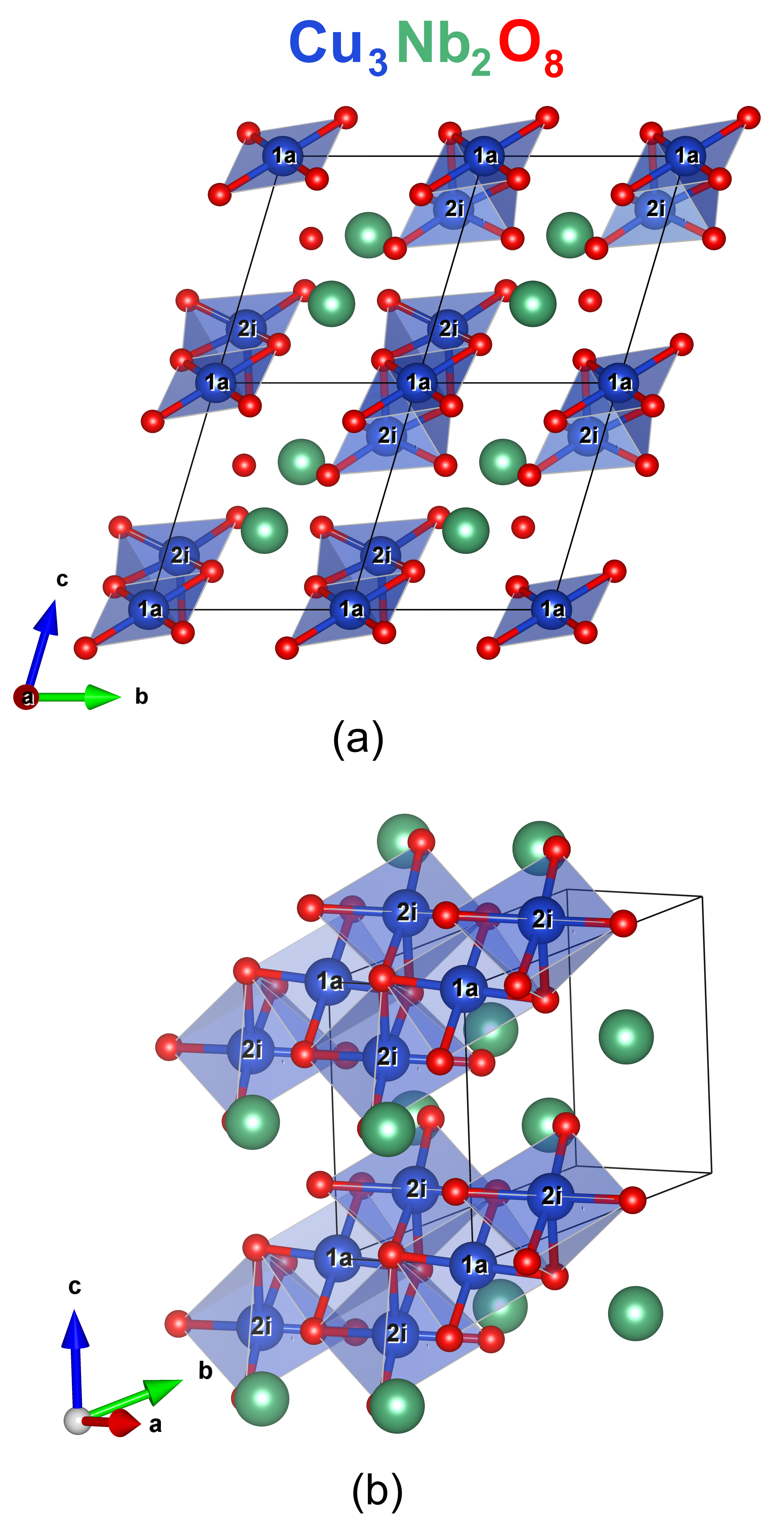}
\caption{\label{fig:NuclStruc} Figure showing the nuclear structure of Cu$_3$Nb$_2$O$_8$. (a) along the a direction where we can see that the Cu sites are separated along b by layers of Nb. (b) shows the saw-tooth chains along a made by the Cu sites. The different Wyckoff sites occupied by Cu are labeled. Figure made in \textsc{VESTA}.\cite{VESTA_Paper}}
\end{figure}

The nuclear structure (as reported by Johnson \textit{et al.}) is shown in Fig. \ref{fig:NuclStruc} with the two Cu Wyckoff sites $1a$ and $2i$ labeled. The $1a$ site has square-planar oxygen coordination while the $2i$ has a square-pyramidal coordination. The structure can be thought of as layers of Cu separated by layers of Nb along the $b$ axis and the Cu sites form saw-tooth chains along the $a$ axis.

Johnson \textit{et al.} reported that the low temperature phase is generically helicoidal with all spins rotating in a common plane and the $1a$ site wholly out of phase with the two $2i$ sites which are slightly out of phase with each other. We use the term \textit{generic helicoid} as an intermediary between the cases of a cycloid (where the propagation vector is contained within the rotation plane) and a helix (where the propagation vector is perpendicular to the rotation plane).

Ferroelectricity is closely linked to helicoidal order as both require noncentrosymmetry.\cite{Spaldin_Review} Thus chiral magnetic ordering can induce ferroelectricity even if the parent phase forbids it by symmetry such as Cr$_2$BeO$_4$\cite{Newnham_CrBeO} or TbMnO$_3$.\cite{Kimura_TMO} The classic model of cycloidal/helicoidal multiferroics is the KNB spin-current model\cite{KNB_Paper} where there exists a strict constraint on the direction of any spontaneous polarization. Where all spins rotate in a common plane, the induced polarization is $\vec{P} \propto \vec{v}_{ij} \times \vec{S}_i \times \vec{S}_j$ where $\vec{v}_{ij}$ is the vector that joins the two neighboring spins $\vec{S}_i$, $\vec{S}_j$ and is also in the rotation plane. This results in a polarization which is constrained to lie in the rotation plane. The KNB model has been verified for many systems such as MnWO$_4$.\cite{MNWO_Paper}

The low temperature polarization observed in Cu$_3$Nb$_2$O$_8$ is reported to be almost perpendicular to the rotation plane.\cite{Johnson_CNO_Paper} This is clearly incompatible with the KNB model and Johnson \textit{et al.} proposed the phenomenological `ferro-axial' model which couples the polarization through a chiral term to a macroscopic axial vector allowed in certain crystals classes by symmetry. In $P\bar{1}$, there is no specified direction of this axial vector and so the polarization may be along an arbitrary direction. 

This model was supported by Sharma \textit{et al.}\cite{CNOSharma2015} However, Xiang \textit{et al.} proposed a more general model of helical multiferroics\cite{WhangboGeneralPaper} in which a polarization is induced purely through the presence of a noncollinear magnetic structure. The polarization results from noncollinear spin dimers (the exchange pairs) and is expanded as a power series and the coefficients determined by first principles DFT calculations. This model was found to explain the polarization displayed by MnI$_2$\cite{WhangboGeneralPaper} and later extended to CaMn$_7$O$_{12}$\cite{WhangboCMOPaper} where a noncentrosymmetric  structure was considered. This model has also been applied to Cu$_3$Nb$_2$O$_8$\cite{WhangboCNOPaper} where it was concluded that the polarization arises from exchange striction between Cu pairs not spin-orbit coupling which they claim is in contradiction with the ferro-axial coupling mechanism. Furthermore, they suggested that the small magnitude of the polarization is due to the small phase difference between the Cu $2i$ sites.

Given this, a conclusive method to classify the magnetic structure is required. In this study, we report the magnetic structure in a single crystal sample of Cu$_3$Nb$_2$O$_8$ as determined by spherical neutron polarimetry (SNP).   There are two goals to this work.  First, we aim to confirm the low temperature magnetic structure in single crystals given the discussion surrounding the mechanism for ferroelectricity.  Second, we aim to investigate the unusual spin density wave phase and its relation to the two magnetic transitions.  We will accomplish this by reporting the magnetic structure in \emph{both} ordered phases using SNP's sensitivity to individual components of the magnetic interaction vector. SNP involves measuring the polarization matrix which contains information about how the sample interacts with the polarized neutron beam. Without the use of polarized neutrons, it is very difficult to distinguish between complex structures such as helicoidal or a spin density wave.

This paper is divided into four sections including this introduction.  To ensure consistency and also to explain the measured polarization matrix elements,  in Section \ref{Pol_Scat} we outline the theoretical background behind SNP, derive the Blume-Maleev equations using the density matrix formalism and describe the instrument setup. Full refinements and temperature dependence of the polarization matrix are presented in Section \ref{Res_Disc} along with the associated discussion. This is followed by Section \ref{Conc} which presents the conclusions of the paper.

\section{Polarized neutron scattering}\label{Pol_Scat}

The scattering of a polarized beam of neutrons with a single crystal is governed by the Blume-Maleev equations.\cite{Blume1963, MaleevPaper} Whereas unpolarized scattering only probes the magnitude of the interaction vector $|\vec{M}_{\perp}|^2$, SNP is directly sensitive to its components. This allows a much greater level of accuracy and affords the ability to distinguish between structures which would appear similar in an unpolarized study.\cite{PJB_Pol1, PJB_Pol2, PJB_Pol3} This has been shown in the case of FeAs\cite{ChrisFeAS} (and also using polarized x-ray scattering)\cite{FeAs2} and CeRhIn$_5$\cite{ChrisCRI, CRI2} where the structures were found to be spin density wave and helical arrangement respectively. Furthermore, it is also for this reason that SNP can determine complex magnetic structures as shown in the cases of CaBa(Co$_3$Fe)O$_7$\cite{CBCFO} and Mn$_2$GeO$_4$.\cite{M2GO4}

In order to unambiguously determine the magnetic structure of Cu$_3$Nb$_2$O$_8$, we first review the Blume-Maleev equations which are presented in Section \ref{BMDer}. This re-derivation condenses the development of this topic  using the density matrix formalism and the properties of Pauli matrices. A comparison of these equations to the data will show several problems that require an evaluation of the systematic errors given in Appendix A.

\subsection{Blume-Maleev equations}\label{BMDer}

State scattering $\ket{\chi^I} \rightarrow \ket{\chi^F}$ can be expressed as a transformation in spin-half space by the  2 $\times$ 2 matrix $S = N + \vec{M}_{\perp} \cdot \vec{\sigma}$: $\ket{\chi^F}=S \ket{\chi^I}$ where $\{\sigma_i\}$ are the Pauli matrices, $N$ corresponds to nuclear scattering, and $\vec{M}_{\perp} \cdot \vec{\sigma}$ is due to magnetic scattering.\cite{Blume1963} We can discount scattering from nuclear spins as these are taken to be disordered and so any linear terms must average to zero.

The scattering cross-section $d \sigma$ is given by the ratio of the number of particles ($N d \Omega$) scattered into the solid angle $d \Omega$ in angular direction $(\theta , \phi)$ per unit time to the incident flux ($|\vec{j}^I|$):\cite{XSecRef}

\begin{equation}
    \frac{d\sigma}{d\Omega} = \frac{N}{|\vec{j}^I|}.
\end{equation}

\noindent Away from the direction of the incident beam ($\hat{k}$) we can rewrite $N d \Omega = \vec{j}^F \cdot d\vec{A}$ where $\vec{j}^F$ is the resultant flux which is number of particles scattered into the solid angle $d \Omega$ in angular direction $(\theta , \phi)$ per unit time per unit area. At a distance $r$, $d\vec{A}$ is given by $r^2 \hat{r} d\Omega$ and 

\begin{equation}\label{eq:crossSecJ}
    \frac{d\sigma}{d\Omega} = \frac{1}{|\vec{j}^I|} \vec{j}^F \cdot \hat{r} r^2.
\end{equation}

\noindent In our reactor-based neutron experiments, the incident flux is a free particle current $\vec{j}^I = \frac{\hbar \vec{k}}{m}$. $\vec{j}^{F}$ is derived from the Schr\"{o}dinger equation:\cite{QuantFlux}

\begin{equation}\label{eq:j}
    \vec{j} =  \frac{\hbar}{m} \mathcal{I}m \Big \{ \psi (\vec{r})^{\dagger} \vec{\nabla} \psi (\vec{r}) \Big \},
\end{equation}

\noindent where $\psi (\vec{r})$ is the total wavefunction projected into real space. Given the wavefunction is a product of a spatial $\ket{\phi}$ and a spin part $\ket{\chi}$: $\ket{\psi} = \ket{\phi} \otimes \ket{\chi}$, the total wavefunction is a linear combination of the inital plus final states

\begin{equation}
	\ket{\psi} = \ket{\phi^I} \otimes \ket{\chi ^I} + \ket{\phi^F} \otimes \ket{\chi ^F}.
\end{equation}

\noindent Considering the spacial part, we project this into coordinate space by left multiplication of $\bra{\vec{r}}$. Then, the initial and final parts are given according to the Born approximation: the projected spacial part of the initial wavefunction is a plane wave $e ^{i \vec{k} \cdot \vec{r}}$ and the final part is a spherical wave $\frac{e ^{ikr}}{r}$:

\begin{equation}
	\begin{split}
		\psi (\vec{r}) = \braket{\vec{r}}{\psi} = & \bra{\vec{r}} \big( \ket{\phi^I} \otimes \ket{\chi ^I} + \ket{\phi^F} \otimes \ket{\chi ^F} \big) \\
	    	                                                        = & e ^{i \vec{k} \cdot \vec{r}} \ket{\chi ^I}  + \frac{e ^{ikr}}{r} \ket{\chi ^F}.
	\end{split}
\end{equation}

Substituting this into Equation \ref{eq:j} and neglecting cross terms, which average to zero, 

\begin{equation}
   	 \vec{j}  = \frac{\hbar \vec{k}}{m} + \frac{\hbar k}{m} \frac{\braket{\chi ^F}}{r^2} \hat{r}.
\end{equation}

\noindent Note we have neglected higher order terms in $\frac{1}{r}$. The first term corresponds to the initial flux $\vec{j}^I$ and therefore the second term must correspond to the outgoing flux $\vec{j}^F$. Substituting these expressions for the flux into Equation \ref{eq:crossSecJ}, the cross-section is

\begin{equation}
	\begin{split}\label{eq:crossSec}
		\frac{d\sigma}{d\Omega} &= \frac{1}{|\frac{\hbar \vec{k}}{m}|} \frac{\hbar k}{m} \frac{\braket{\chi ^F}}{r^2} r^2 \\
		                                       &= \braket{\chi ^F} = \mel{\chi ^I}{S^\dagger S}{\chi ^I} \\
		                                       &= Tr(\rho S^\dagger S),
   	\end{split}
\end{equation}

\noindent where $\rho = \dyad{\chi ^I}$ is the density matrix.  This result is also expected from Fermi's golden rule. Using $\{\sigma_i\}$ along with the identity ($\mathbb{I}$) as a basis to span the space of $2 \times 2$ Hermitian matrices (of which $\rho$ belongs), we can express  $\rho$ (using Einstein summation notation)

\begin{equation}
	\rho = a \mathbb{I} + b_i \sigma _i
\end{equation}

\noindent and, noting the following relations for the traces of Pauli matrices

\begin{equation}\label{eq:Pauli}
\begin{split}
	Tr( \sigma_i) & = 0, \\
   	Tr( \sigma_i \sigma_j) &= 2 \delta_{ij}, \\
	Tr(\sigma_i  \sigma_j  \sigma_k)&=2i\epsilon_{ijk}, \\
	Tr(\sigma_i \sigma_j \sigma_k \sigma_l)&=2(\delta_{ij}\delta_{kl} - \delta_{ik}\delta_{jl} + \delta_{il}\delta_{jk}),
\end{split}
\end{equation}

\noindent we can determine $a$ and $b_i$ which are real as $\rho$ is Hermitian.  To determine the coefficient $a$, we consider the trace of $\rho$:

\begin{equation}
	\begin{split}
		Tr(\rho) = & Tr( a \mathbb{I} + b_i \sigma _i) \\ 
			     = & a Tr(\mathbb{I}) + b_i Tr(\sigma _i) \\
			     = & a Tr(\mathbb{I}) \\
			     = & 2a.
	\end{split}
\end{equation}

\noindent However, we can also write this (using an arbitrary basis $\{ \ket{u_i} \}$) as:

\begin{equation}
	\begin{split}
		Tr(\rho) = & Tr(\dyad{\chi ^I}) \\
			     = & \braket{u_i}{\chi^I} \braket{\chi^I}{u_i} \\ 
			     = & \braket{\chi^I}{u_i} \braket{u_i}{\chi^I} \\ 
			     = & \braket{\chi^I}{\chi^I} = 1,
	\end{split}
\end{equation}

\noindent where we have used closure or completeness of the basis $\{ \ket{u_i} \}$ ($\ket{u_i} \bra{u_i} = \mathbb{I}$) and the fact that $\ket{\chi^I}$ is normalized. We therefore have

\begin{equation}
	a = \frac{1}{2}.
\end{equation}

We find $b_i$ by considering the polarization which is defined as an average of  $\vec{\sigma}$: $P_i^I = \expval{\sigma_i}$. Using the density matrix to perform, we find

\begin{equation}
	\begin{split}
		P_i^I = \expval{\sigma_i} = & Tr(\rho \sigma_i) \\
		       				       = & Tr( [\frac{1}{2} \mathbb{I} + b_j \sigma _j] \sigma_i) \\
						       = & \frac{1}{2} Tr(\sigma_i) + b_j Tr( \sigma _j \sigma_i) \\
						       = & 2 b_j \delta_{ij} = 2b_i.
	\end{split}
\end{equation}

Using the expressions for $a,b$, can write $\rho$ in terms of the incident polarization:

\begin{equation}\label{eq:denMat}
	\rho = \frac{1}{2}(\mathbb{I} + \vec{P}^I \cdot \vec{\sigma}).
\end{equation}

\noindent This dependence of the density matrix makes sense physically. If, for example, the system is fully polarized in the `up' direction, then all the wavefunction density will be concentrated into the `up' state and this is reflected in $\rho$. Similarly, if the state is unpolarized then we find $\rho$ is diagonal.

An expression for the final polarization can be constructed as an average of $\vec{\sigma}$ over the \emph{final} spin state:

\begin{equation}
       P_i^{F} = \frac{\expval{\sigma_i}{\chi^F}}{\braket{\chi^F}} = \frac{Tr( \rho S^\dagger \sigma_i S)}{\frac{d\sigma}{d\Omega}}.
\end{equation}
 
\noindent Importantly, as $S$ is not unitary, the norm of a state is \emph{not} conserved in the scattering process and so this average must be normalized appropriately. 

By computing these traces (using the properties of the Pauli matrices \textit{viz.} Equation \ref{eq:Pauli}), we can calculate the cross-section and final polarization. These are the Blume-Maleev equations:\cite{Blume1963, MaleevPaper}

\begin{widetext}
	\begin{equation} \label{eqn:XSec} 
	        \frac{d\sigma}{d\Omega} = Tr(S \rho S^\dagger) = |N|^2 + | \vec{M}_{\perp}|^2 + N(P^I_i  M_{\perp i}^*) + N^*(P^I_i  M_{\perp i}) - i\epsilon_{ijk} P^I_i (M_{\perp j} M_{\perp k}^*),
	\end{equation}

	\begin{equation} \label{eq:BM2}
		\begin{split}
	       P_i^F = \frac{1}{\frac{d\sigma}{d\Omega}} Tr(S \rho S^\dagger \sigma_i) = & \frac{1}{\frac{d\sigma}{d\Omega}} \Big{[}(|N|^2  - |\vec{M}_{\perp}|^2)\delta_{ij} +  i( N^* M_{\perp k}- N M_{\perp k} ^* ) \epsilon_{ijk} + M_{\perp i} M_{\perp j}^* + M_{\perp j} M_{\perp i}^* \Big{]} P^I_j +\hdots \\
																 & +\frac{1}{\frac{d\sigma}{d\Omega}} \Big{[} N M_{\perp i}^* +  N^* M_{\perp i} + i (\vec{M}_{\perp} \times \vec{M}_{\perp}^*)_i \Big{]}\\
														 	\equiv & \mathsf{P}_{ij}P^I_j + P^{\prime}_i,
		\end{split}
	\end{equation}
\end{widetext}

\noindent where $\mathsf{P}$ is the `polarization tensor' and contain information about how the polarization rotates during the scattering. $\vec{P}^{\prime}$ corresponds to the polarization created during the interaction.

\begin{figure*}[t!]
	\includegraphics[width=.95\linewidth]{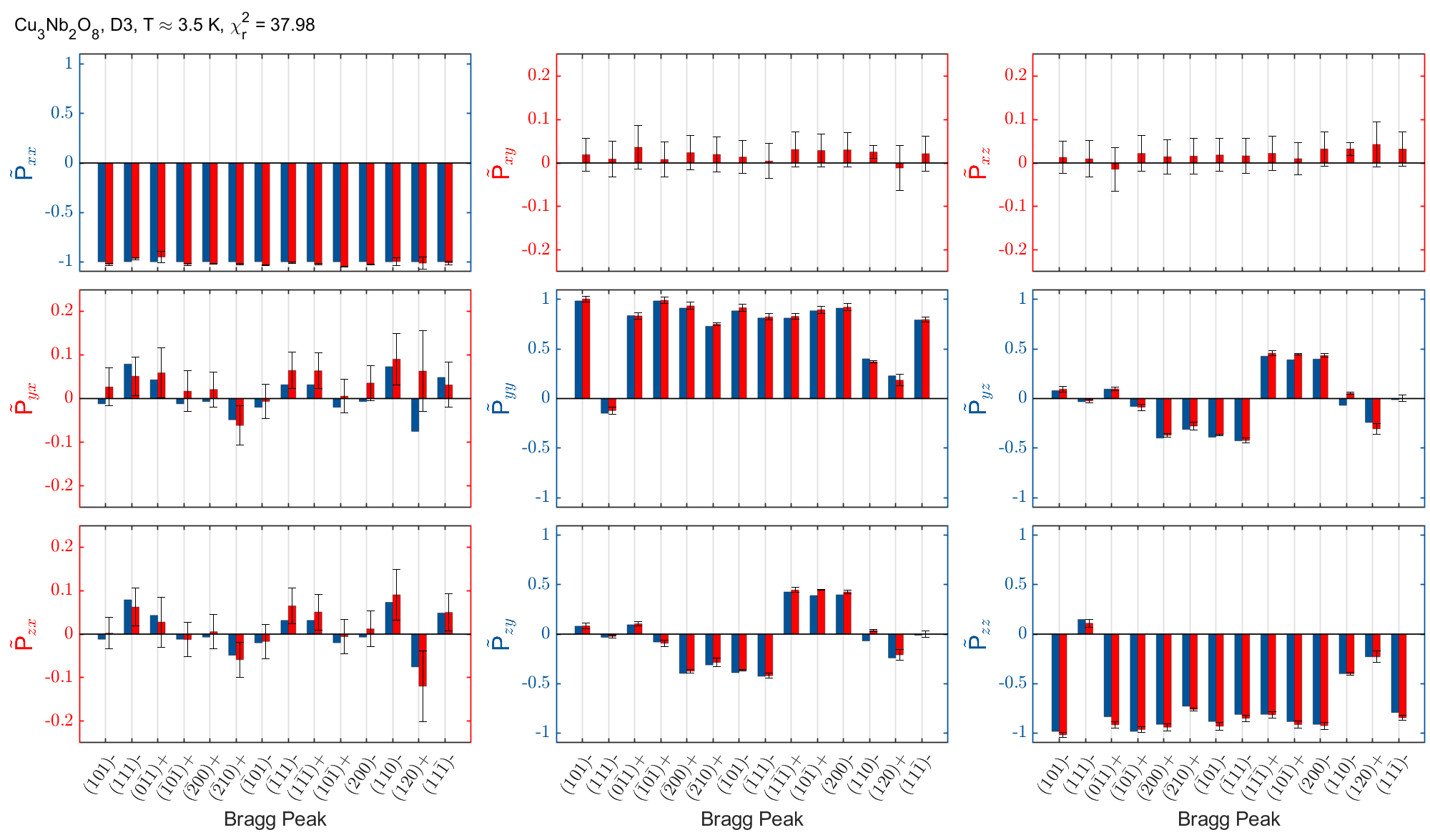}
\caption{\label{fig:FitsL}Figure showing the refinement (in \textsc{Mag2Pol} and confirmed in MATLAB) of the polarization matrix at $\approx 3.5 K$ -  LT phase. The bars show the refined matrix elements (left - blue in color) plotted for each magnetic Bragg peak against the measured matrix elements (right - red in color). Statistical and systematic experimental errors are shown in black (systematic errors due to instrument resolution were calculated in \textsc{MATLAB} and are discussed in more detail in Appendix A). The plotted matrix elements are corrected for detector spin filter efficiency. For clarity, two different $y$ scales are used and are displayed in different colors. Magnetic Bragg peaks are labelled as $(hkl)\pm$ meaning $(hkl) \pm \vec{k}$. We note that the systematic deviations of $\tilde{\mathsf{P}}_{xx}$ from -1 and also $\tilde{\mathsf{P}}_{xy}$ and $\tilde{\mathsf{P}}_{xz}$ from 0 are discussed in the main text.}
\end{figure*}

\begin{figure*}[t!]
	\includegraphics[width=.95\linewidth]{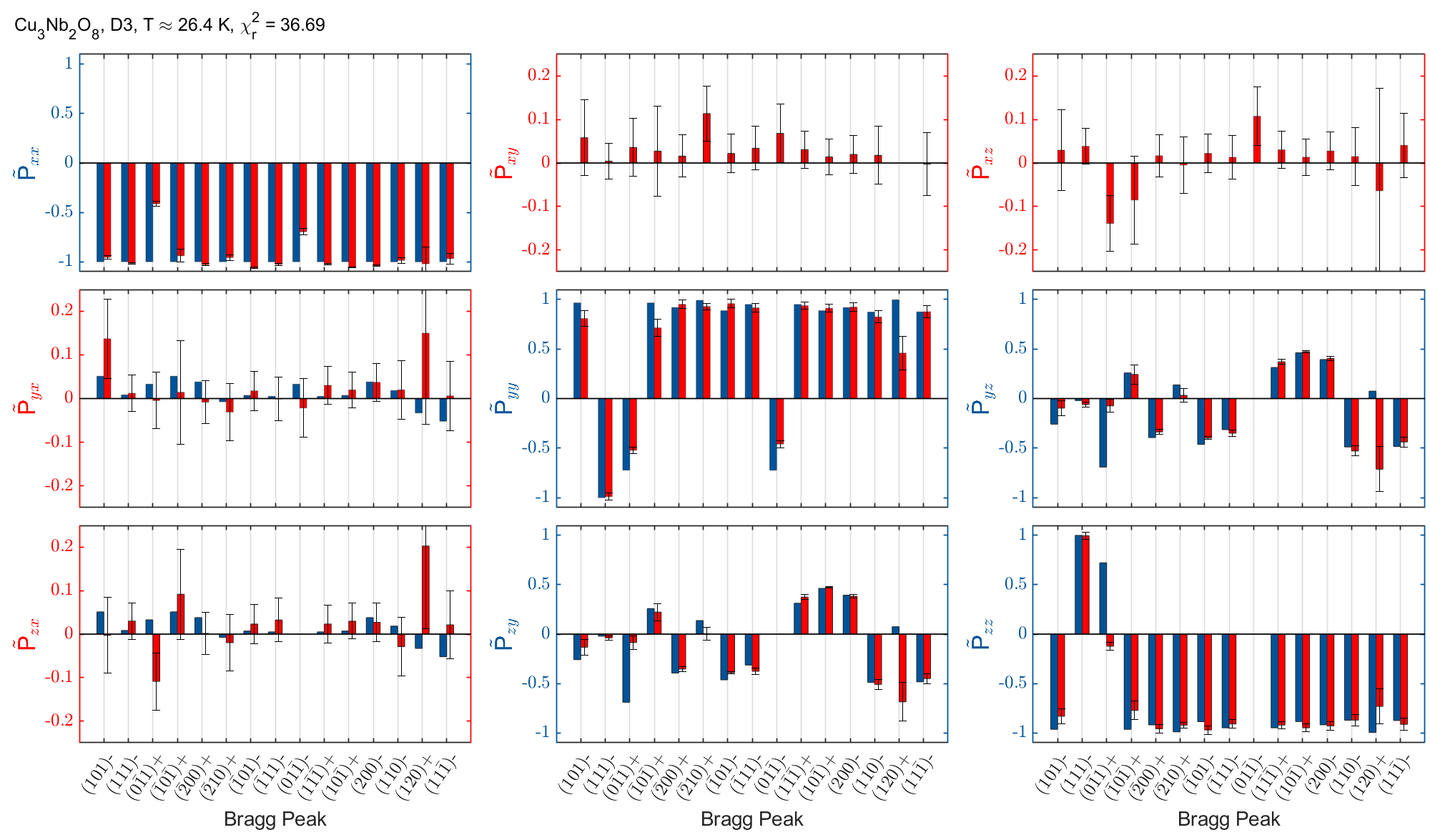}
\caption{\label{fig:FitsM}Figure showing the refinement (in \textsc{Mag2Pol} in MATLAB) of the polarization matrix at $\approx 26.4 K$ -  MT phase. The bars show the refined matrix elements (left - blue in color) plotted for each magnetic Bragg peak against the measured matrix elements (right - red in color). Statistical and systematic experimental errors are shown in black (systematic errors due to instrument resolution were calculated in \textsc{MATLAB} and are discussed in more detail in Appendix A). The plotted matrix elements are corrected for detector spin filter efficiency. For clarity, two different $y$ scales are used and are displayed in different colors. Magnetic Bragg peaks are labelled as $(hkl)\pm$ meaning $(hkl) \pm \vec{k}$.  We note that the systematic deviations of $\tilde{\mathsf{P}}_{xx}$ from -1 and also $\tilde{\mathsf{P}}_{xy}$ and $\tilde{\mathsf{P}}_{xz}$ from 0 are discussed in the main text.}
\end{figure*}

When a crystal has a non-zero propagation vector (like here in Cu$_{3}$Nb$_{2}$O$_{8}$), magnetic Bragg peaks occur as satellites of the nuclear ones. These will be referred to here (Figs. \ref{fig:FitsL} and \ref{fig:FitsM}) as 

\begin{equation}
	(hkl)\pm \equiv (hkl) \pm \vec{k},
\end{equation}

\noindent for the remainder of the paper. For Cu$_3$Nb$_2$O$_8$, the propagation vector is $\vec{k} = (0.4876, 0.2813, 0.2029)$ and therefore we set $N = 0$ (no nuclear contribution) in equations \ref{eqn:XSec} and \ref{eq:BM2}. 

\subsection{Experimental method}\label{exp_Meth}
The `polarization matrix', with components $\tilde{\mathsf{P}}_{ij}$, is defined as the ratio of scattered polarization in the $j$th direction to an incident polarization which is in the $i$th direction. It is conventional to use the so-called `standard' co-ordinates with $x \parallel$ to the scattering vector, $z$ vertical and $y$ completing a right-handed co-ordinate system.  $\tilde{\mathsf{P}}$ can be defined using Equations \ref{eqn:XSec} and \ref{eq:BM2} as

\begin{equation}\label{eq:polMat0}
	\tilde{\mathsf{P}}_{ij} = \Bigg{\langle} \frac{\mathsf{P}_{jk}P^I_k + P^{\prime}_j}{P^I_i} \Bigg{\rangle},
\end{equation}

\noindent where the angled brackets indicate an average over domains (it is important to also note that the cross-section contained within $\mathsf{P}$ and $\vec{P}^{\prime}$ must also be averaged over domains).\cite{PJBrown_Book} However, if we assume a fully polarized initial beam along one axis this can be rewritten as (with \emph{no} Einstein convention):

\begin{equation}\label{eq:polMat}
	\tilde{\mathsf{P}}_{ij} = \Big{\langle} \mathsf{P}_{ji} + P^{\prime}_j \Big{\rangle}.
\end{equation}

Now, it is useful to consider $\frac{d\sigma}{d\Omega}$, $\mathsf{P}$ and $\vec{P}^{\prime}$ in the `standard' co-ordinates only as, in this basis, the $x$ component of $ \vec{M}_{\perp}$ is \emph{always} zero:\cite{PJBrown_Book}

\begin{equation} 
        \frac{d\sigma}{d\Omega}=  |\vec{M}_{\perp}|^2 + 2P^I_x \mathcal{I}m \{ M_{\perp y} M_{\perp z}^* \},
\end{equation}

\begin{equation}
	\mathsf{P} = \frac{1}{\frac{d\sigma}{d\Omega}} \begin{pmatrix}
	-|\vec{M}_{\perp}|^2 & 0 & 0 \\
	0 & |M_{\perp y}|^2 - |M_{\perp z}|^2 & 2 \mathcal{R}e \{ M_{\perp y} M_{\perp z}^* \} \\
	0 & 2 \mathcal{R}e \{ M_{\perp y} M_{\perp z}^* \} & |M_{\perp z}|^2 	- |M_{\perp y}|^2 	
	\end{pmatrix},
\end{equation}

\begin{equation}
	\vec{P}^{\prime} = \frac{1}{\frac{d\sigma}{d\Omega}} \begin{pmatrix}
	-2\mathcal{I}m \{ M_{\perp y} M_{\perp z}^* \} \\
	0 \\
	0
	\end{pmatrix}.
\end{equation}

\noindent Substituting these into Equation \ref{eq:polMat} for a single domain structure gives

\begin{equation} \label{eq:polMat2}
	\tilde{\mathsf{P}} = \frac{1}{|\vec{M}_{\perp}|^2} \begin{pmatrix}
	-|\vec{M}_{\perp}|^2 & 0 & 0 \\
	A & B & C \\
	A & C & -B 	
	\end{pmatrix},
\end{equation}

\noindent where $A = - 2 \mathcal{I}m \{ M_{\perp y} M_{\perp z}^* \}$, $B = |M_{\perp y}|^2 - |M_{\perp z}|^2$ and $C = 2 \mathcal{R}e \{ M_{\perp y} M_{\perp z}^* \}$.  

There are two key points to note about this matrix for the purposes of our analysis.  First, the $\tilde{\mathsf{P}}_{xx} \equiv -1$ is required for magnetic scattering.  Second, $\tilde{\mathsf{P}}_{xy}=\tilde{\mathsf{P}}_{xz} \equiv 0$.  This point will be discussed below in the context of the experimentally measured matrix elements motivating an analysis of the errors in SNP.

Naturally, this expression must be averaged over all domains if present. If the structure is chiral, it may have multiple domains of opposite chirality with each domain polarizing the beam in the opposite way due to the opposing handednesses. This has the results that $\tilde{\mathsf{P}}_{yx}$ and $\tilde{\mathsf{P}}_{zx}$ will cancel out under equal chiral domains. This can also be seen mathematically as these two terms are resultant from a cross-product which is odd under a chiral inversion. Practically, this effect can be offset by, for example, cooling under an electric field to offset the domain population.\cite{NavidPolEField} 

Experimentally, all components of the polarization matrix can be directly measured using CRYOPAD (Cryogenic Polarization Analysis Device).\cite{CryopadPaper}  Developed at the ILL, CRYOPAD is a method of performing SNP. It consists of a cryostat surrounded by two cylindrical Meissner shields with superconducting coils in between. The Meissner shields ensure the sample space is field free. The coils along with incoming and outgoing nutators, allow the polarization vector of the neutron beam to be orientated in any direction. The up-down orientation of the scattered neutron beam is measured using a $^3$He detector.\cite{PJBrown_Book} In this way, the number of neutrons aligned with ($n_j^{(+)}$) or against ($n_j^{(-)}$) the desired measurement axis ($j$) can be measured for any given initial polarization direction ($i$). This gives a means of connecting the intensity with the polarization matrix

\begin{equation}
	\tilde{\mathsf{P}}_{ij} = \frac{(n_j^{(+)} - n_j^{(-)})}{(n_j^{(+)} + n_j^{(-)})},
\end{equation}

\noindent where we have assumed that the initial beam is fully polarized.  This relation provides a direct link between the measured intensity and the polarization matrix elements.  Throughout the rest of the paper we only quote the matrix elements $\tilde{\mathsf{P}}_{ij}$.

\section{Results and discussion}\label{Res_Disc}

We now apply the theory outlined above to understand the neutron polarization results from Cu$_{3}$Nb$_{2}$O$_{8}$.  We first describe the refinement and reconcile some of the apparent inconsistencies between the data and the Blume-Maleev equations.  We then apply this to refine the magnetic structure.

\subsection{Full matrix refinement and application of systematic errors}

\begin{figure}[t!]
	\includegraphics[width=.8\linewidth]{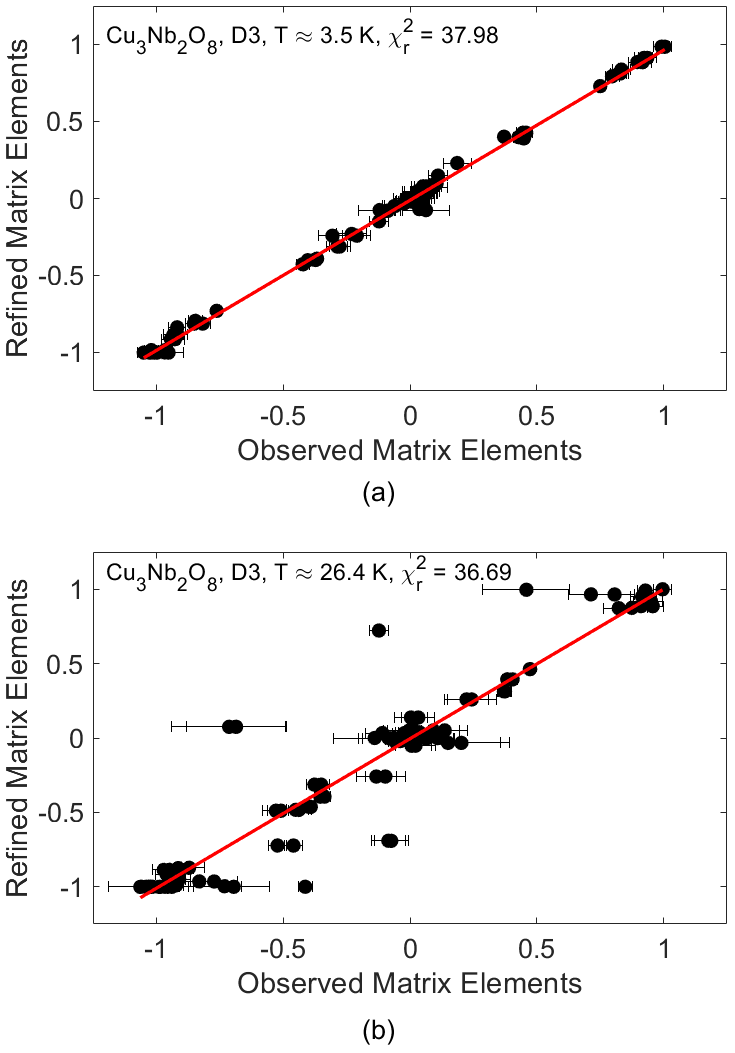}
\caption{\label{fig:ObsCal}Figure showing the refined (in Mag2Pol) elements of the polarization matrix against observed in (a) $\approx 3.5 K$ -  LT phase and (b) $\approx 26.4 K$ -  MT phase. The error bars shown are the systematic (as discussed in Appendix A) and statistical errors on the observed elements.}
\end{figure}

In order to determine the magnetic structure of Cu$_3$Nb$_2$O$_8$ in both ordered phases, SNP was used to measure the polarization matrix for multiple magnetic Bragg peaks on a single crystal grown by the floating zone technique. The experiment was performed using CRYOPAD on D3 (ILL, Grenoble; $\lambda = 0.85 \text{ \normalfont\AA}$) which allows the polarization matrix to be directly measured (as detailed in Section \ref{exp_Meth}). The full matrix was determined for multiple magnetic Bragg peaks at $\approx 3.5K$ (LT phase) and just below $T_N$ at $\approx 26.4K$ (MT phase). These temperatures are both deep in the two phases proposed by Johnson \textit{et al.} The data-sets were refined in \textsc{Mag2Pol}\cite{Mag2PolRef} and cross checked using the Blume-Maleev equations in MATLAB.   Fig. \ref{fig:FitsL} and Fig. \ref{fig:FitsM} show the outcome of the refinement in the LT and MT phase respectively. Also shown in Fig. \ref{fig:ObsCal} are the refined matrix elements plotted against their measured counterparts for both phases. The overlaid linear pattern provides an indication that the refinement is of good quality. 

The three Cu$^{2+}$ ions occupy the Wyckoff positions $1a$ and $2i$; the latter are identical in all but the LT phase where the inversion centre is broken. This symmetry was taken into account in the MT phase by setting the moments of the Cu$(2i)$ sites to be identical. Constraints regarding the lengths of the moments were also implemented into \textsc{Mag2Pol} refinement as SNP is not sensitive to the lengths of the moments unless there is nuclear-magnetic overlap.\cite{PJBrown_Book} This means that SNP cannot be used to determine the absolute length of the magnetic moments, only their relative size and directions. However, this usually provides sufficient information to ascertain the magnetic ground state, especially given published work refining the magnetic moment value~\cite{Johnson_CNO_Paper}.

Given the Blume-Maleev equations discussed above, there were two strict rules governing the polarization matrix: the elements $\tilde{\mathsf{P}}_{xx}\equiv -1$ and $\tilde{\mathsf{P}}_{xy}=\tilde{\mathsf{P}}_{xz}=0$.  From the matrix elements plotted in Figs. \ref{fig:FitsL} and Fig. \ref{fig:FitsM}, these two results do not seem to be obeyed with the $\tilde{\mathsf{P}}_{xx}$ consistently $\neq -1$ and the elements  $\tilde{\mathsf{P}}_{xy}=\tilde{\mathsf{P}}_{xz}$, though small (with the $y$-axis highlighted by red) $\neq$ 0.  We now address these two points in detail.

Due to the chiral nature of the LT structure, the option to refine using uncorrected matrix elements was implemented in \textsc{Mag2Pol} (see Qureshi).\cite{Mag2PolRef} The detector on D3 relies on a $^3$He spin filter which, over time, will decay. This results in a reduced intensity of measurements\cite{Mag2PolRef} but can be corrected for in \textsc{Mag2Pol} and so was taken into account for the refinement process. However, the correction process is computed only during the refinement due to the presence of terms which do and don't depend on the initial polarization. In order to quantify the goodness of the fit the reduced $\chi ^2$ on the uncorrected matrix elements was calculated by \textsc{Mag2Pol}\cite{Mag2PolRef} as $\chi_r^2 = 37.98$ and $\chi_r^2 = 36.69$ in the LT and MT phases respectively. The data were later corrected for spin filter efficiency and Fig. \ref{fig:FitsL} and Fig. \ref{fig:FitsM} show the \emph{corrected} matrix elements.  It is the correction resulting from the incomplete beam polarization that is the dominant reason for the matrix element $\tilde{\mathsf{P}}_{xx}$ deviating from -1.

Finally, regarding the refinement, we must address the effects of the experimental setup to understand the systematic $\neq$ 0 values for the matrix elements $\tilde{\mathsf{P}}_{xy}$  and $\tilde{\mathsf{P}}_{xz}$ . We have already discussed the detector but, furthermore, the initial polarization of the neutron beam may not be perfect. This can be due to two reasons: the monochromator used is not 100\% effective or the magnetic fields used to align the beam's polarization to the desired direction have a finite resolution (which may also be due to sample alignment). Both of these effects are purely due to instrumental/sample alignment precision and are discussed more fully in Appendix A. However, the non-zero values for $\tilde{\mathsf{P}}_{xy}$ and $\tilde{\mathsf{P}}_{xz}$ shown in Fig. \ref{fig:FitsL} and Fig. \ref{fig:FitsM} can be fully accounted for by a beam polarization that is misaligned (corresponding to the angular resolution of 2$^{\circ}$ quoted in the literature\cite{CryopadPaper} which reduces the beam polarization by 0.06\%). As such we expect that actual systematic experimental errors should be considered larger than the statistical counting errors. The polarization is known to be 93.5\% on D3 so we can assume that these effects will have a non-negligible impact on our results. These errors were computed using custom code implemented in \textsc{MATLAB}. Due to the nature of these errors, individual components of $\vec{M}_{\perp}$ were required to be calculated and used according to the equations given in Appendix A. The systematic errors are plotted in Fig. \ref{fig:FitsL} and Fig. \ref{fig:FitsM} (but were not taken into account in the calculation of $\chi_r^2$). 

Having understood the matrix elements and deviations from some of the strict rules established by the Blume-Maleev equations, we now discuss the results for the refined magnetic structures.

\subsection{Refined magnetic structure}\label{MagStrucSec}

\begin{figure}[h!]
		\includegraphics[width=.7\linewidth]{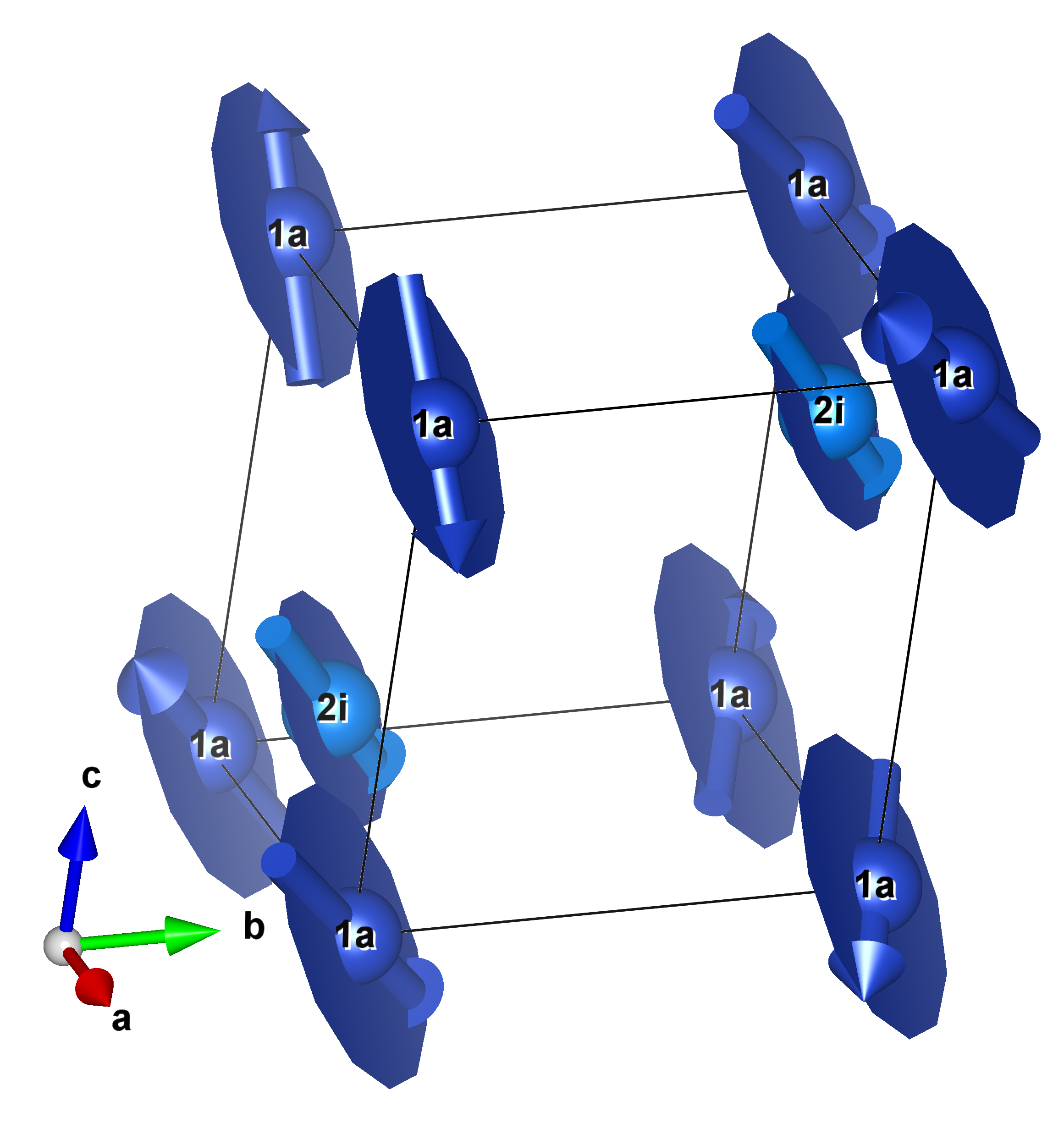}
		\caption{\label{fig:LTMagStruc} Figure showing the refined magnetic structure of Cu$_3$Nb$_2$O$_8$ in the LT phase. The Cu$^{2+}$ magnetic moments are shown along with their rotational envelope. The two Wyckoff positions are labeled and shown in a different shade of blue for clarity. Figure made in \textsc{VESTA}.\cite{VESTA_Paper}}
	\end{figure}

\begin{figure}[h!]
	\includegraphics[width=.95\linewidth]{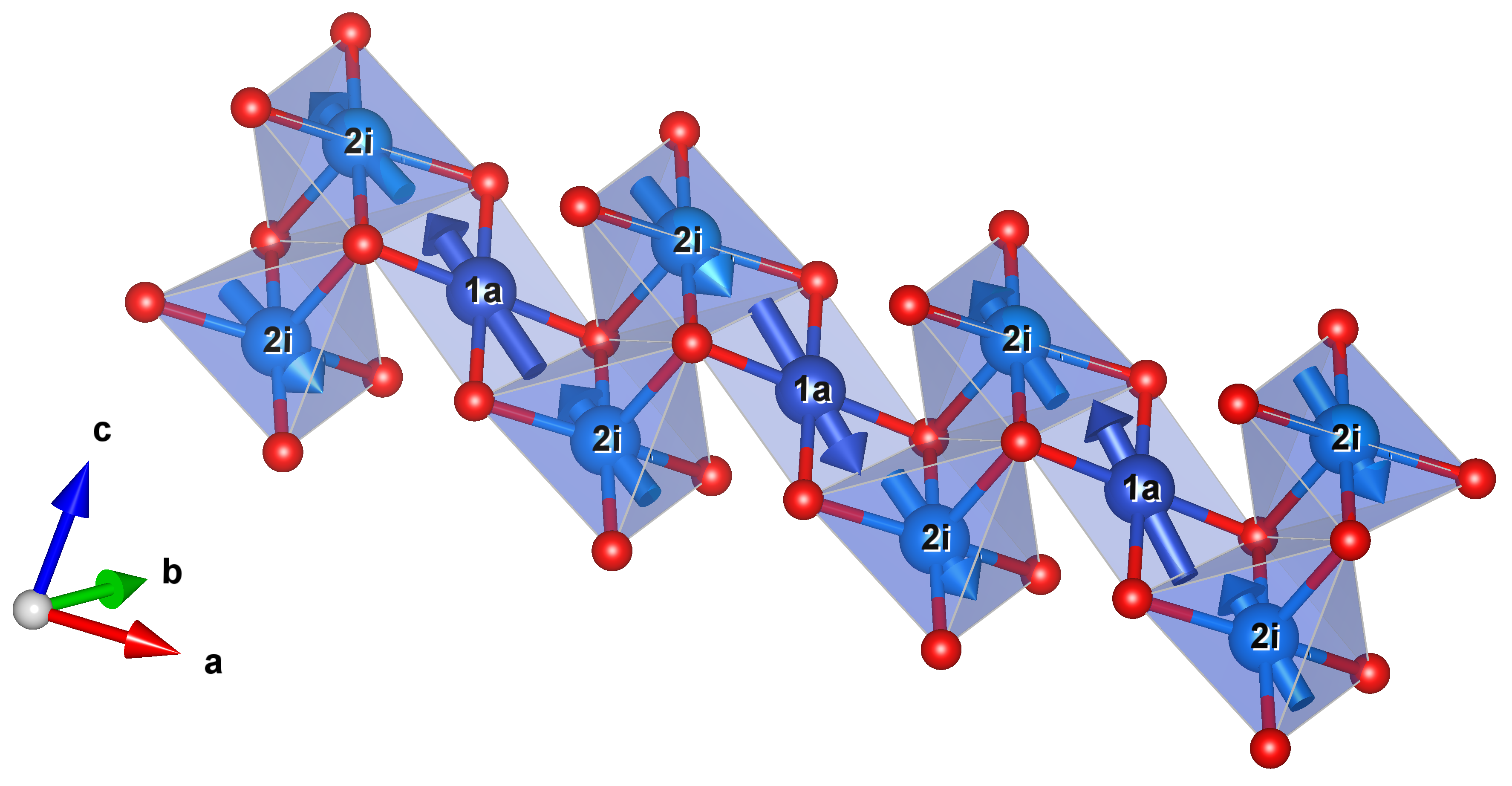}
	\caption{\label{fig:CuChain} Figure showing the Cu trimer saw-tooth chain. These trimers are ferromagnetically aligned but antiferromagentically aligned with neighboring trimers. The two Wyckoff positions are labeled and shown in a different shade of blue for clarity. Figure made in \textsc{VESTA}.\cite{VESTA_Paper}}
\end{figure}

\begin{figure}[hb!]
	\includegraphics[width=.95\linewidth]{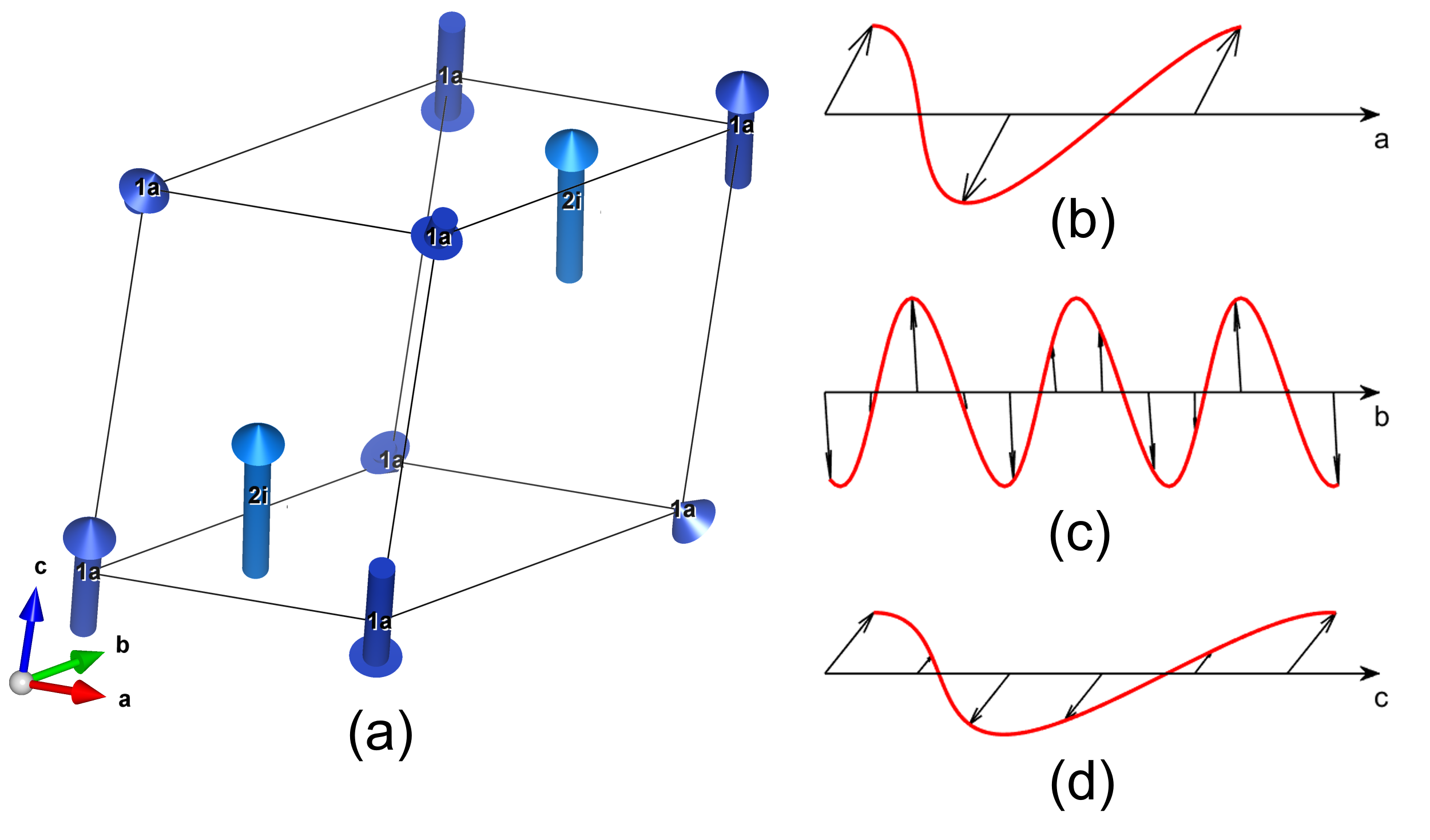}
	\caption{\label{fig:MTMagStruc} Figure showing the refined magnetic structure of Cu$_3$Nb$_2$O$_8$ in the MT phase. The Cu$^{2+}$ magnetic moments in the context of the unit cell is shown in (a) whilst the in-plane oscillations along each crystal axis are shown in (b) - (d). As we can approximate the propagation vector as ($\frac{1}{2}$, $\frac{3}{11}$, $\frac{1}{5}$) and so hence we should expect approximately one complete oscillation along $a$ within two unit cells, three complete oscillations along $b$ within eleven unit cells and one complete oscillation along $c$ within five unit cells as is seen in the figure. The out-of-plane oscillations are much smaller in comparison and so are not plotted. This is due to $\mathcal{I}m\{M_{\perp}(\vec{Q})\} \ll \mathcal{R}e\{M_{\perp}(\vec{Q})\}$ in this phase. The two Wyckoff positions are labeled and shown in a different shade of blue for clarity. Figure (a) made in \textsc{VESTA}.\cite{VESTA_Paper}}
\end{figure}

The magnetic structures that result from the above refinements are detailed in Table \ref{tab:LowTStruc} and Table \ref{tab:MidTStruc}, and illustrated in Fig. \ref{fig:LTMagStruc}, Fig. \ref{fig:CuChain}, and Fig. \ref{fig:MTMagStruc}. They can both be described by a rotating spin model given by:

\begin{equation} \label{eqn:Spin}
        S_i(\vec{L})= \mathcal{R}_i \cos(\vec{k} \cdot \vec{L} + \Phi_i) + \mathcal{I}_i \sin(\vec{k} \cdot \vec{L} +  \Phi_i),
\end{equation}

\noindent where $i$ labels the Cu sites and $\vec{L}$ is a real space lattice vector. Coordinates in this section are given with respect to a spherical polar coordinate system $(r,\phi,\theta)$ constructed inside an orthonormal basis $(x^{\prime}y^{\prime}z^{\prime})$ where $x^{\prime} \parallel a$ with $b$ in the $x^{\prime}-y^{\prime}$ plane.

In the LT phase, the ground state of the system exhibits a generic helicoidal structure (Table \ref{tab:LowTStruc}, Fig. \ref{fig:LTMagStruc}).  Here the spin rotation is confined to the plane spanned by the real and imaginary parts of $\vec{M}_{\perp}(\vec{Q})$. This plane can be wholly described by its normal whose direction may be written in angular coordinates $(\phi,\theta)$. In this study, we find a rotation plane whose normal has angular coordinates $(\phi,\theta)$ to be $(59.73^{\circ},80.81^{\circ})$ for the Cu$(1a)$ site and $(59.78^{\circ},81.00^{\circ})$ for Cu$(2i)$.  As noted above, CRYOPAD is not sensitive to the absolute moment value, but as shown in Table \ref{tab:LowTStruc}, the ratio of the Cu$^{2+}$ moments on different sites is consistent with the unpolarized work giving a value of $|M_{Cu1}/M_{Cu2,3}|=0.89 \mu_{B} /0.69 \mu_{B}$.~\cite{Johnson_CNO_Paper}

In this phase, the Cu$^{2+}$ sites form ferromagnetic trimers which in turn form an antiferromagnetic saw-tooth chain along the $a$ direction. This is illustrated in Figure \ref{fig:CuChain}. Two chiral domains (as reflected by the loss of the inversion centre) were permitted in the refinement and the populations were refined to be roughly equal (46\%\textbackslash54\%). This accounts for the small value of $\tilde{\mathsf{P}}_{yx}$ and $\tilde{\mathsf{P}}_{zx}$ as these terms have the opposite sign in the two domains as discussed above in the context of the Blume-Maleev equations. 

This structure is broadly in agreement with Johnson \textit{et al.} (plane normal of $(54.9^{\circ},75.5^{\circ})$\cite{Johnson_CNO_Paper} gives a discrepancy of $\approx 7^{\circ}$). However, whereas Johnson \textit{et al.} reported a helicoidal structure with a circular rotation envelope from their powder sample, the best refinement of the SNP data resulted in an elliptical envelope. In the refinement presented in Fig. \ref{fig:FitsL}, the length of the imaginary part of $M(\vec{Q})$ is $76\%$ of the real part. This results in an elliptical rotation envelope with eccentricity of $0.65$ compared to $0$ from the circular structure reported by Johnson \textit{et al.} A refinement was attempted which included a circular constraint but it produced a worse fit to the data (included in Appendix B). However, this experiment was performed at finite temperature and it is possible the circular envelope is recovered $T \rightarrow 0$.

In the structure reported here, the electric polarization is still out of the rotation plane at $\approx 17^{\circ}$ to the plane normal. Moreover, although the phase difference ($\approx \pi$) between the Cu$(1a)$ and Cu$(2i)$ sites strongly agrees with that reported by Johnson \textit{et al.}, it is worth noting that the different coordinate systems used result in an overall $\approx 55 ^{\circ}$ phase factor.

\begin{table}[h!]
\setlength{\tabcolsep}{0.5em} 
\caption{\label{tab:LowTStruc} \centering Table showing the refined magnetic structure at $\approx 3.5 K$. Magnitudes are normalized.}
\begin{tabular}{c c c c c}\hline \hline 
	& $|M|$ (arb. units) & $\phi$ ($^{\circ}$) & $\theta$ ($^{\circ}$) & $\Phi$ ($2 \pi$ rad.) \\ [0.5ex] \hline
	$\mathcal{R}_1$ & $1$  & $-43.62$ & $35.01$ & $0$ \\
	$\mathcal{I}_1$ & $ 0.76$  & $-23.93$ & $124.32$ &  \\
	$\mathcal{R}_2$ & $0.78$  & $-43.41$ & $34.77$ & $0.515$ \\
	$\mathcal{I}_2$ & $0.59$  & $-24.04$ & $123.15$ & \\
	$\mathcal{R}_3$ & $0.78$  & $-43.41$ & $34.77$ & $0.525$ \\
	$\mathcal{I}_3$ & $0.59$  & $-24.04$ & $123.15$ & \\ [1ex] \hline
\end{tabular}
\end{table}

\begin{table}[h!]
\setlength{\tabcolsep}{0.5em} 
\caption{\label{tab:MidTStruc} \centering Table showing the refined magnetic structure at $\approx 26.4 K$. Due to symmetry ($P\bar{1}$) $\mathcal{R}_2 = \mathcal{R}_3$ and $\mathcal{I}_2 = \mathcal{I}_3$ and so the latter are omitted from the table. Magnitudes are normalized.}
\begin{tabular}{c c c c c}\hline \hline 
	& $|M|$ (arb. units) & $\phi$ ($^{\circ}$) & $\theta$ ($^{\circ}$) & $\Phi$ ($2 \pi$ rad.) \\ [0.5ex] \hline
	$\mathcal{R}_1$ & $0.75$  & $-43.00$ & $30.60$ & $0$ \\
	$\mathcal{I}_1$ & $0.063$  & $-154.36$ & $80.46$ &  \\
	$\mathcal{R}_2$ & $1$  & $-46.35$ & $30.34$ & $0.000$ \\
	$\mathcal{I}_2$ & $0.036$  & $-150.28$ & $72.99$ & \\ [1ex] \hline
\end{tabular}
\end{table}

In the MT phase, a spin density wave (SDW) structure is refined as the ground state (Table \ref{tab:MidTStruc}, Fig. \ref{fig:MTMagStruc}). As in the LT case, the spin can be described by the rotating model with the rotation plane given by the real and imaginary parts of $M(\vec{Q})$. An SDW results when one of these become small compared to the other. In the refined structure, $\mathcal{I}m\{M_{\perp}(\vec{Q})\}$ becomes almost zero resulting in a highly elliptical rotational envelope which manifests as a modulation of the spins (see Fig. \ref{fig:MTMagStruc} $(b)$ - $(c)$). The polarization of this SDW coincides with the LT rotation plane to within $2.5^{\circ}$. Also, all Cu$^{2+}$ sites are now in phase. With the loss of time reversal symmetry at $T_N$ we should expect 180$^{\circ}$ domains to be present in this phase. However, as these will produce the same polarization matrix, only one domain was included in the refinement.

\subsection{Temperature dependence}

\begin{figure}[h!]
	\includegraphics[width=.95\linewidth]{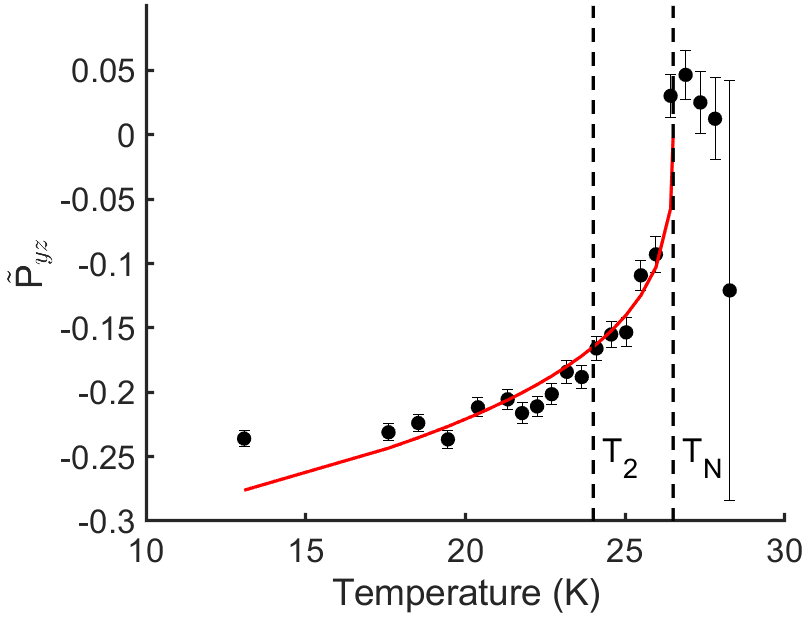}
	\caption{\label{fig:VsT}Plot of matrix element $\tilde{\mathsf{P}}_{yz}$ against temperature. This was measured on the Bragg peak $(\bar{2}10)+$. The N\'{e}el temperature $T_N \approx 26.5 K$ is indicated. The fit (solid line) shows a power law $|T-T_N|^{2\beta}$ with exponent $\beta = 0.154$.}
\end{figure}

The temperature dependence of the polarization matrix was also measured. Figure \ref{fig:VsT} shows the element $\tilde{\mathsf{P}}_{yz}$  measured on the Bragg peak $(\bar{2}10)+$ against temperature. A power law $|T - T_N|^{2\beta} $ was fitted with an exponent $\beta = 0.154$. The fact that this is centered on $T_N$ reflects that $\tilde{\mathsf{P}}_{yz}$ is indicative of magnetic ordering. This exponent is consistent with an Ising interpretation.\cite{IsingRef} This can be motivated by considering the crystal field as the O polyhedra surrounding the Cu$^{2+}$ are distorted by the crystal field which induces an anisotropy in a direction in which there is an energy cost associated with flipping a spin.~\cite{Yosida}

In the remainder of this paper we shall discuss the possible mechanisms behind the reported magnetic structure and, in particular, the presence of a SDW in the MT phase. Triclinic Cu$_{3}$Nb$_{3}$O$_{8}$ is constrained by few symmetry elements with only an inversion center in the paramagnetic phase and, as such, the magnetic free energy near T$_{N}$ can therefore be expanded in symmetry allowed even powers of the components of the site magnetization $\vec{M}$\cite{TolAndTol} as Equation \ref{eq:F1}. For the purpose of this simple illustration we may neglect any cross and gradient terms. Owing to the lack of symmetry elements with only the presence of an inversion center, there is no requirement that the $\alpha_{i}$ components are equal and a magnetic transition occurs when one of these goes to zero. However, even with the assumption that the magnitude of $\vec{M}$ must be fixed, we can write this as Equation \ref{eq:F2} which preserves the fact that each component of the order parameter may have different temperature dependencies.   

\begin{widetext}
	\begin{equation}
	f_M(T)=f_{0}+\alpha_{x}(T)|M_{x}|^{2}+\alpha_{y}(T)|M_{y}|^{2}+\alpha_{z}(T)|M_{z}|^{2} +  \sum_{i,j}\beta_{ij}|M_{i}|^{2}|M_{j}|^{2} + \cdots \label{eq:F1}
	\end{equation}
	\begin{equation}
	f_M(T)=f_{0}+(\alpha_{x}(T)-\alpha_{z}(T))|M_{x}|^{2}+(\alpha_{y}(T)-\alpha_{z}(T))|M_{y}|^{2}+\alpha_{z}(T)|M|^{2} + \sum_{i,j}\beta_{ij}|M_{i}|^{2}|M_{j}|^{2} + \cdots \label{eq:F2}
	\end{equation}
\end{widetext}

This description of the phase transition gives an anisotropy in the spatial structure of the real space magnetism near the magnetic transition. However, it is not a spin density wave in the context of what is observed in metallic systems owing to a nesting wave vector across an electronic Fermi surface.\cite{SDWRef1,SDWRef2}  This is corroborated by our measurement of the temperature dependence of the off diagonal term in the polarization matrix $\tilde{\mathsf{P}}_{yz}$ which shows little response to the second transition. Furthermore, if the structure did become a collinear spin density wave in this phase, we should expect all off diagonal terms in the polarization matrix to be zero - this is not observed in Fig. \ref{fig:VsT}. We therefore conclude that the ``spin density wave phase'' in Cu$_{3}$Nb$_{2}$O$_{8}$ near the Neel transition is rather a manifestation of the symmetry allowed decoupling of the different components of the order parameter.

In this way the decoupled magnetic structure destabilizes the crystal structure to the point where it induces the structural chirality at $T_2$ due to the presence of critical fluctuations around  T$_{N}$. This implies that the two transitions are indirectly coupled in an analogue with the Jahn-Teller effect\cite{JTRef} ({\it c.f.} MgV$_2$O$_4$, ZnV$_2$O$_4$).\cite{MVO1,MVO2,MVO3,ZVO} However, in that case, structural distortion to lift the orbital degeneracy lowers the symmetry and this allows magnetic ordering to occur at a lower temperature. 

Microscopically this mechanism can be motivated from the inverse Dzyaloshinskii-Moriya effect (whereby a magnetic structure with a helical component will induce structural chirality) and is compatible with ferro-axial coupling\cite{Johnson_CNO_Paper} and other symmetry considerations.\cite{Hlinka1,Hlinka2} The generic helicoidal structure is returned at low temperatures and is consistent with our measurements. Choosing the $z$ direction to be component that goes to zero at T$_{N}$,  such that $\alpha_{z}(T)\propto |T-T_{N}|$, at a temperature below T$_{N}$ one component of the magnetization will dominate the free energy.  Terms in the free energy coupling magnetization and structural order parameters will then allow it to become energetically favorable for the structure to distort as observed in Cu$_{3}$Nb$_{2}$O$_{8}$ when the transition from P$\overline{1}$ $\rightarrow$ P1 occurs.  

 Further analogy may be drawn with the nematic phase in Fe-based superconductors (i.e. the Fe pnictides).\cite{FeSC_Rev} Here it is argued the introduction of one type of ordering induces (via symmetry) others and thus the nematic order must be considered as resultant from `correlation-driven electronic instabilities', which are likely driven by magnetic fluctuations. Similarly, in the case of Fe$_{1+x}$Te, a spin density wave structure is observed near the phase transition which is reported to be stabilized by magnetic fluctuations \cite{FeTe_1,FeTe_2}.

We speculate that many of the `SDW' phases reported in the literature for magnetic insulators maybe due to the decoupling of the different components of the order parameter. In analogy with a fictitious force, whilst these SDW phases appear genuine, they are simply the result of a deeper mechanism at work. Measuring the temperature dependence of the polarization matrix is clearly important in understanding these transitions. Indeed, in the case of Ni$_3$V$_2$O$_8$\cite{NVO_1,NVO_2} a similar mechanism has also been proposed in order to account for the presence of such a magnetic and ferroelectric structure where the ferroelectricity is due to a spin induced symmetry breaking.\cite{NVO_3} This has interesting consequences for the controlling of electric properties in these materials by applied magnetic fields which merits further study.

\section{Conclusions}\label{Conc}

In conclusion, spherical neutron polarimetry was used to study the magnetic structure of Cu$_3$Nb$_2$O$_8$. The full polarization matrix was determined in both low temperature phases for multiple magnetic Bragg peaks and the structure was refined to an apparent spin density wave below $T_N \approx 26.5K$, which becomes generically helicoidal below $\approx 24K$. The low temperature phase was found to be generally in agreement with the powder structure reported by Johnson \textit{et al}. The temperature dependence of the matrix was also measured and the critical exponent extracted. We propose a mechanism which could explain the presence of the SDW in this insulator. The structure, which manifests as an imitation of a SDW at finite temperatures, is actually reflective of the symmetry allowed decoupling of the components of the order parameter allowing one to dominate the free energy. In turn, this then allows (through a coupling between magnetic and structural order parameters) the structural distortion and the manifestation of the electric polarization.

\section*{Acknowledgments}

The authors would like to thank H. Lane for helpful discussions. We are grateful for funding from the EPSRC, ILL and the STFC.  NGD supported by EPSRC/Thales UK iCASE Award EP/P510506. Work at Rutgers University was supported by the DOE under Grant No. DOE: DE-FG02-07ER46382. The work at Postech was supported by the National Research Foundation of Korea(NRF) funded by the Ministry of Science and ICT(No. 2016K1A4A4A01922028)

\setcounter{equation}{0}
\renewcommand{\theequation}{A\arabic{equation}}
\section*{Appendix A: Initial polarization Error Analysis}

In a scattering experiment we need to consider not just the statistical errors that arise from the measurement procedure, but also any systematic errors that may be present. This discussion is motivated by the observation that the polarization matrix elements $\tilde{\mathsf{P}}_{xy}$ and $\tilde{\mathsf{P}}_{xz}$ (as shown in Fig. \ref{fig:FitsL} and Fig. \ref{fig:FitsM}) are non-zero despite being predicted as such by Equation \ref{eq:polMat2}. We can surely attribute this to experimental error but the size of these uncertainties should be quantified. As such, we can attribute all systematic errors to two mechanisms:

\begin{enumerate}
\item{Not all of the neutron beam is polarized - only a fraction $\zeta \approx 1$ - the rest remains unpolarized.}
\item{The polarization vector is not exactly aligned to the specified direction (i.e. $x$, $y$, etc.) or some other \emph{small} misalignement is present}
\end{enumerate}

The causes of these mechanisms lay in the instrumental setup: The first case is consequent of a non-ideal monochromator crystal which leaves part of the beam unpolarized. The second is dependent on the apparatus used to align the polarization vector of the neutron beam. On D3 this is a setup of a guide magnetic field and a magnetic nutator. Now this \emph{small} misalignment can be due to either the incoming/outgoing polarization vector or the sample not being aligned correctly. This effects can all be included into an 'angular resolution' parameter $\theta$. We shall take, without loss of generality, the outgoing polarization as exact in this treatment.

We can treat the first mechanism easily using the density matrix formalism. If we have a partially polarized beam then we need to construct a density matrix that represents this `mixed state'. This can be done by combining the density matrices that correspond to the separate `pure states' multiplied by their respective population fraction in the beam:

\begin{equation}
        \rho_{mixed} =  \sum_i n_i \rho_i,
\end{equation}

\noindent where $i$ count the number of pure states that are being combined and the population fractions $\{ n_i \}$ sum to unity. Now the density matrix of the polarized fraction is given by Equation \ref{eq:denMat} so we only need to compute the density matrix of the unpolarized section. This can be done in two ways: either we can just set $\vec{P}^I = 0$ in Equation \ref{eq:denMat} or we can construct this matrix from first principles. We know that an unpolarized beam will be made up of equal parts spin up and down neutrons (where the spin axis is chosen as $z$). We see that an unpolarized beam is also a `mixed state' and so construct is density matrix according to

\begin{equation}
\begin{split}
        \rho_{unpol} =  \frac{1}{2}\rho_{up} +  \frac{1}{2}\rho_{down} = & \frac{1}{2} \begin{pmatrix}1 & 0 \\ 0 & 0\end{pmatrix} + \frac{1}{2} \begin{pmatrix}0 & 0 \\ 0 & 1\end{pmatrix} \\ = & \frac{1}{2} \mathbb{I}.
\end{split}
\end{equation}

\noindent Hence, we can combine this with \ref{eq:denMat} to write the full density matrix

\begin{equation}
\begin{split}
        \rho_{1} = & \zeta \rho_{pol} +  (1 - \zeta) \rho_{unpol} \\
	= & \big [ \frac{\zeta}{2} (\mathbb{I} + P^I_i \sigma_i)\big ] + \frac{(1 - \zeta)}{2} \mathbb{I} \\
	= & \frac{1}{2} (\mathbb{I} + \zeta P^I_i \sigma_i).
\end{split}
\end{equation}

Interestingly, we note that this has the same form as Equation \ref{eq:denMat} except that we acquire a factor of $\zeta$ such that we must now use

\begin{equation}
	\vec{P}^I \rightarrow \zeta \vec{P}^I
\end{equation}

in our analysis. Notice that this can only affect the magnitude of the polarization matrix elements. We must therefore conclude that the anomalous values of $\tilde{\mathsf{P}}_{xy}$ and $\tilde{\mathsf{P}}_{xz}$ must result from the second mechanism, which we will now consider:

Let us suppose that the alignment the polarization vector with the direction $\hat{n}$ has an angular resolution of $\theta$. This means we can construct a cone (with angle $\theta$) around $\hat{n}$ within which we expect the polarization vector to be contained. We shall consider the worse case here and specify that the polarization vector lies on the surface of the cone. In this case we can describe it's direction using two angles: $\theta$ - the angular deviation from $\hat{n}$ - and $\phi$ - the azimuth angle at the base of the cone. We can see that our polarization vector will now, in the general case, acquire components in the other two Cartesian directions perpendicular to $\hat{n}$. So if we allow the initial polarization vector to acquire these extra components, the final polarization in our measured direction will also contain terms which come from the scattering of these other components. Hence, this effect is able to change the form of the polarization matrix as the components will contain these extra contributions. This effect is able to explain why we have observed a non-zero value for $\tilde{\mathsf{P}}_{xy}$ and $\tilde{\mathsf{P}}_{xz}$. As the cause of the misalignment comes from the instrumentation, we can expect that it should retain a consistent value during a measurement (i.e. one matrix element) and will only be reset when either the polarization vector direction or the sample rotation is changed. Hence, we can expect to see a small but non-zero contribution to all matrix elements.

Let us now compute the error that we can expect from these effects. We can write the initial polarization vector as

\begin{equation}\label{eq:init_Err}
	\vec{P}^I \rightarrow \zeta (\vec{P}^I + \vec{\alpha}^{(i)}(\theta, \phi)),
\end{equation}

\noindent where $\vec{\alpha}^{(i)}(\theta, \phi)$ contains the additional contribution from the non-zero angular resolution when the initial polarization is in the $i$th direction. We can then write an equation for the error $\epsilon_{ij}$ in the polarization matrix element $ \tilde{\mathsf{P}}_{ij}$ as

\begin{equation}
	\epsilon_{ij} = \max_{0 < \phi < 2\pi} \Bigg | \Bigg{\langle} \frac{\zeta \mathsf{P}_{jk} [P_k^I + \alpha_k^{(i)}(\theta, \phi)] + P_j^{\prime}}{\zeta \cos(\theta)} \Bigg{\rangle} -  \tilde{\mathsf{P}}_{ij} \Bigg |,
\end{equation} 

\noindent where we must be careful to use expression \ref{eq:init_Err} for the calculation of the cross-section when calculating the components of the polarization tensor $\mathsf{P}_{jk}$ and the created polarization vector $\vec{P}^{\prime}$. Again the angled brackets indicate an average over domains. We must also specify that this error is given by the maximum value of the function inside the absolute value sign. This is because, we cannot know which value of $\phi$ we should take for each measurement - it being attributed randomly due to the experimental precision. Hence, we should take the maximum to give us the 'worst case' error.

Now, the vector $\vec{\alpha}^{(i)}(\theta, \phi)$ will be dependent on the initial polarization direction and we can write it as

\begin{align}
        \vec{\alpha}^{(x)} = & [\cos(\theta) - 1] \hat{x} + \sin(\theta)[\cos(\phi) \hat{y} + \sin(\phi) \hat{z}], \\
        \vec{\alpha}^{(y)} = & [\cos(\theta) - 1] \hat{y} + \sin(\theta)[\cos(\phi) \hat{z} + \sin(\phi) \hat{x}], \\
        \vec{\alpha}^{(z)} = & [\cos(\theta) - 1] \hat{z} + \sin(\theta)[\cos(\phi) \hat{x} + \sin(\phi) \hat{y}],
\end{align}

\noindent for an initial polarization in the $x$, $y$ and $z$ direction respectively (we are still using the standard coordinate system). 

These error terms can then be computed as:

\begin{widetext}
	\begin{align}
	        \epsilon_{xx} = & \Bigg | \Bigg{\langle} \frac{- \zeta |\vec{M}_{\perp}|^2 \cos(\theta) - 2 \mathcal{I}m \{ M_{\perp y} M_{\perp z}^* \}}{\zeta \cos(\theta) \big{\langle} |\vec{M}_{\perp}|^2 + 2 \zeta \cos(\theta) \mathcal{I}m \{ M_{\perp y} M_{\perp z}^* \}\big{\rangle}} \Bigg{\rangle} + 1 \Bigg | \label{eq:errorBegin}, \\
		\epsilon_{xy} = & \max_{0 < \phi < 2\pi} \Bigg | tan(\theta) \frac{\big{\langle} [|\vec{M}_{\perp y}|^2 - |\vec{M}_{\perp z}|^2] \cos(\phi) + 2 \mathcal{R}e \{ M_{\perp y} M_{\perp z}^* \}\sin(\phi) \big{\rangle}}{\big{\langle} |\vec{M}_{\perp}|^2 + 2 \zeta \cos(\theta) \mathcal{I}m \{ M_{\perp y} M_{\perp z}^* \} \big{\rangle}} \Bigg |, \\
		\epsilon_{xz} = & \max_{0 < \phi < 2\pi} \Bigg | tan(\theta)  \frac{\big{\langle} 2 \mathcal{R}e \{ M_{\perp y} M_{\perp z}^* \} \cos(\phi) - [|\vec{M}_{\perp y}|^2 - |\vec{M}_{\perp z}|^2] \sin(\phi)\big{\rangle}}{\big{\langle}|\vec{M}_{\perp}|^2 + 2 \zeta \cos(\theta) \mathcal{I}m \{ M_{\perp y} M_{\perp z}^* \}\big{\rangle}} \Bigg |, \\
		\epsilon_{yx} = & \max_{0 < \phi < 2\pi} \Bigg | \Bigg{\langle} \frac{ - \zeta |\vec{M}_{\perp}|^2 \sin(\theta) \sin(\phi) - 2 \mathcal{I}m \{ M_{\perp y} M_{\perp z}^* \}}{\zeta \cos(\theta) \big{\langle}|\vec{M}_{\perp}|^2 + 2 \zeta \sin(\theta) \sin(\phi) \mathcal{I}m \{ M_{\perp y} M_{\perp z}^* \}\big{\rangle}} \Bigg{\rangle} -  \tilde{\mathsf{P}}_{yx} \Bigg |, \\
		\epsilon_{yy} = & \max_{0 < \phi < 2\pi} \Bigg | \frac{\big{\langle} |\vec{M}_{\perp y}|^2 - |\vec{M}_{\perp z}|^2 + 2 tan(\theta)\cos(\phi)\mathcal{R}e \{ M_{\perp y} M_{\perp z}^* \}\big{\rangle}}{\big{\langle}|\vec{M}_{\perp}|^2 + 2 \zeta \sin(\theta) \sin(\phi) \mathcal{I}m \{ M_{\perp y} M_{\perp z}^* \}\big{\rangle}}  -  \tilde{\mathsf{P}}_{yy} \Bigg |, \\
		\epsilon_{yz} = & \max_{0 < \phi < 2\pi} \Bigg | \frac{\big{\langle} 2 \mathcal{R}e \{ M_{\perp y} M_{\perp z}^* \} - [|\vec{M}_{\perp y}|^2 - |\vec{M}_{\perp z}|^2] tan(\theta)\cos(\phi)\big{\rangle}}{\big{\langle}|\vec{M}_{\perp}|^2 + 2 \zeta \sin(\theta) \sin(\phi) \mathcal{I}m \{ M_{\perp y} M_{\perp z}^* \}\big{\rangle}} -  \tilde{\mathsf{P}}_{yz} \Bigg |, \\
		\epsilon_{zx} = & \max_{0 < \phi < 2\pi} \Bigg | \Bigg{\langle} \frac{ - \zeta |\vec{M}_{\perp}|^2 \sin(\theta) \cos(\phi) - 2 \mathcal{I}m \{ M_{\perp y} M_{\perp z}^* \}}{\zeta \cos(\theta) \big{\langle}|\vec{M}_{\perp}|^2 + 2 \zeta \sin(\theta) \cos(\phi) \mathcal{I}m \{ M_{\perp y} M_{\perp z}^* \}\big{\rangle}} \Bigg{\rangle} -  \tilde{\mathsf{P}}_{zx} \Bigg |, \\
		\epsilon_{zy} = & \max_{0 < \phi < 2\pi} \Bigg | \frac{\big{\langle} tan(\theta)\sin(\phi)[|\vec{M}_{\perp y}|^2 - |\vec{M}_{\perp z}|^2] + 2 \mathcal{R}e \{ M_{\perp y} M_{\perp z}^* \}\big{\rangle}}{\big{\langle}|\vec{M}_{\perp}|^2 + 2 \zeta \sin(\theta) \cos(\phi) \mathcal{I}m \{ M_{\perp y} M_{\perp z}^* \}\big{\rangle}} -  \tilde{\mathsf{P}}_{zy} \Bigg |, \\
		\epsilon_{zz} = & \max_{0 < \phi < 2\pi} \Bigg | \frac{\big{\langle} 2 tan(\theta)\sin(\phi) \mathcal{R}e \{ M_{\perp y} M_{\perp z}^* \} - [|\vec{M}_{\perp y}|^2 - |\vec{M}_{\perp z}|^2]\big{\rangle}}{\big{\langle}|\vec{M}_{\perp}|^2 + 2 \zeta \sin(\theta) \cos(\phi) \mathcal{I}m \{ M_{\perp y} M_{\perp z}^* \}\big{\rangle}} -  \tilde{\mathsf{P}}_{zz} \Bigg |. \label{eq:errorEnd}
	\end{align}
\end{widetext}

\begin{figure*}
	\centering
	\includegraphics[width=.95\linewidth]{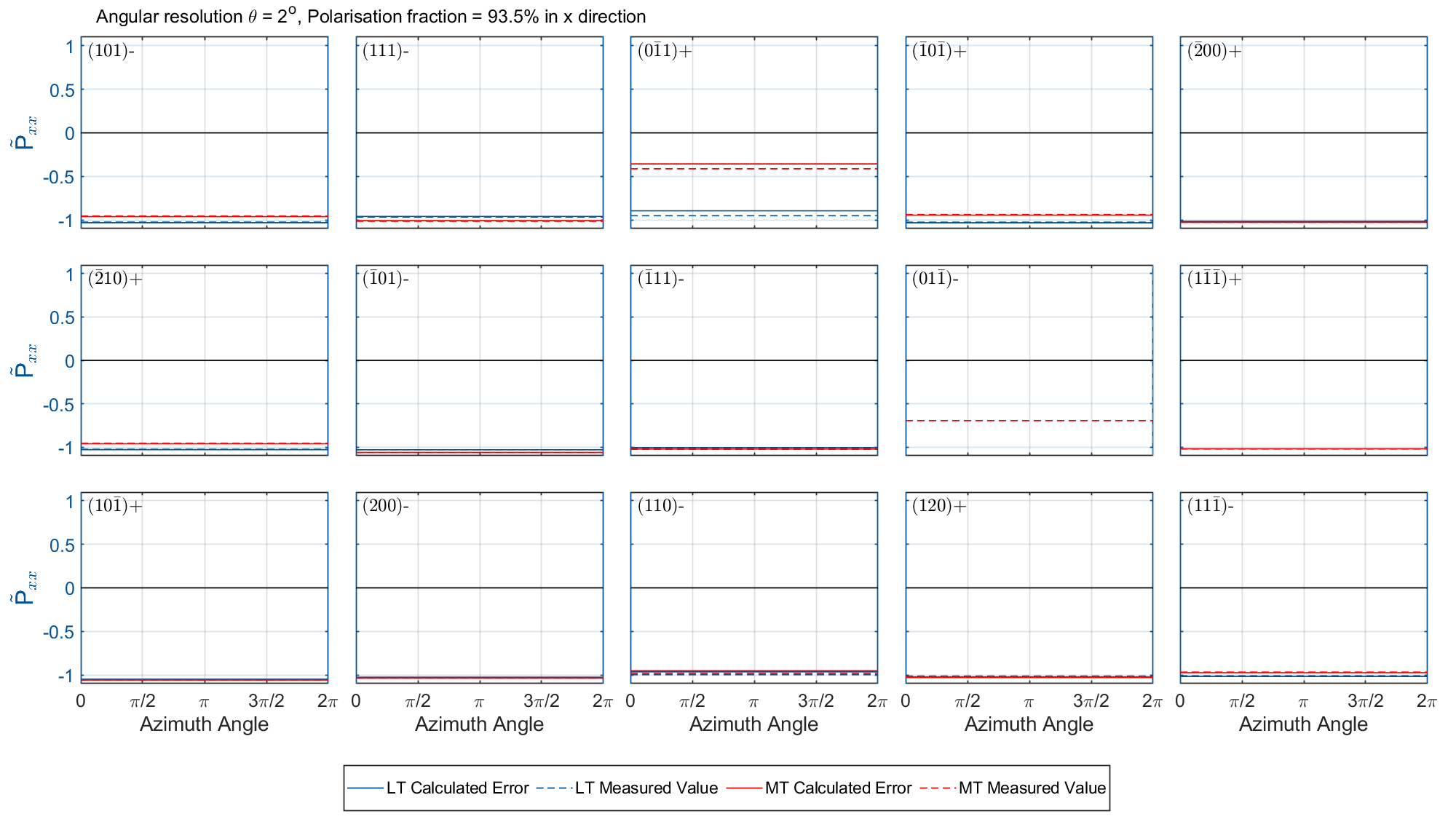}
\caption{\label{fig:PxxError} Figure showing the calculated value of $\tilde{\mathsf{P}}_{xx}$ when an angular resolution of 2$^{\circ}$ is assumed on the incident neutron polarization. The 15 Bragg peaks considered in this study are included an labeled. The solid curves indicate $\tilde{\mathsf{P}}_{xx}$ against the azimuth angle $\phi$ so that the amplitude of these curves give us the `worst case' value for the error. The dashed lines show the measured value of $\tilde{\mathsf{P}}_{xx}$ for each Bragg peak. The two phases are shown in different colors: LT is blue and MT is red.}
\end{figure*}

\begin{figure*}
	\centering
	\includegraphics[width=.95\linewidth]{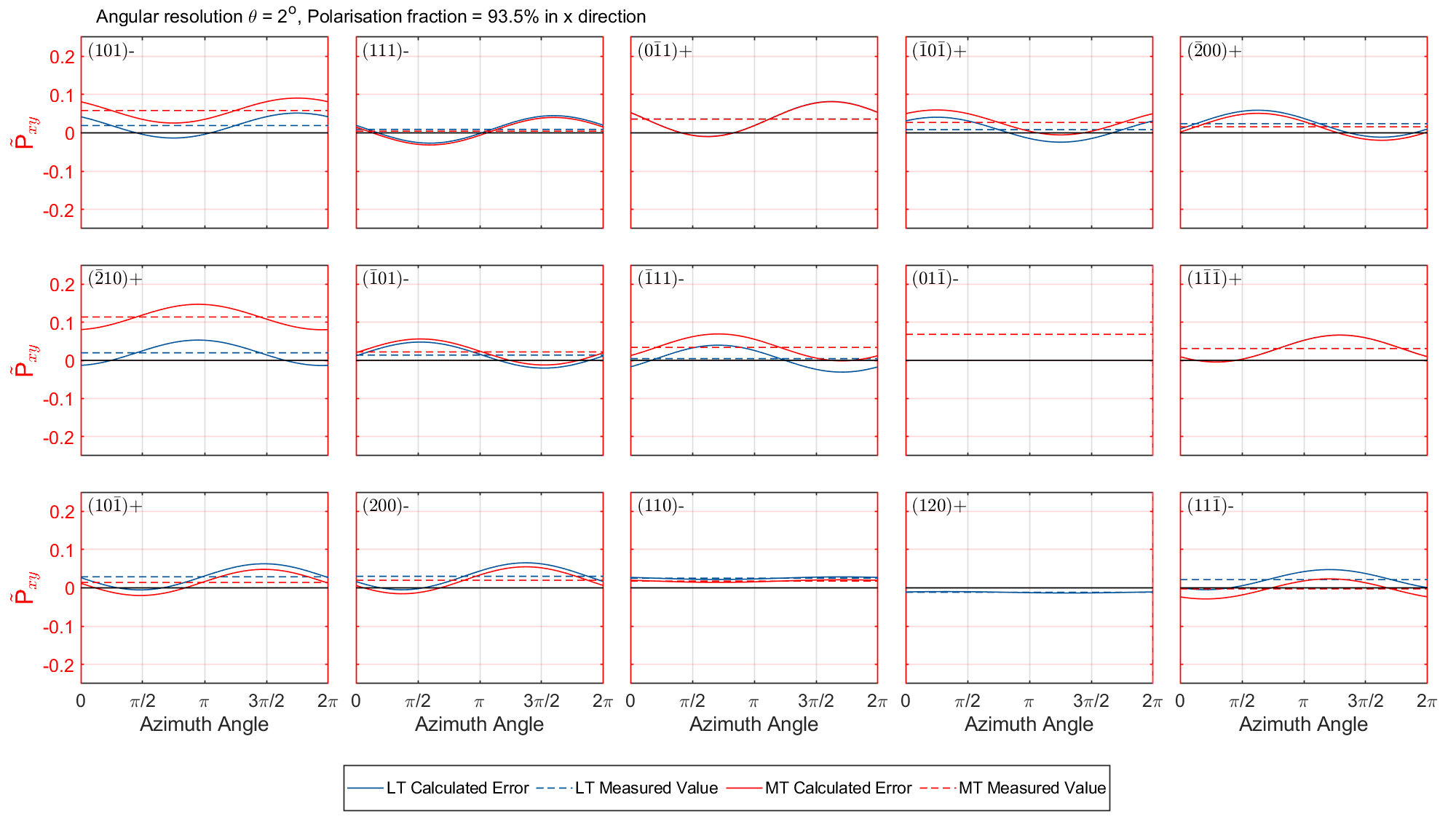}
\caption{\label{fig:PxyError} Figure showing the calculated value of $\tilde{\mathsf{P}}_{xy}$ when an angular resolution of 2$^{\circ}$ is assumed on the incident neutron polarization. The 15 Bragg peaks considered in this study are included an labeled. The solid curves indicate $\tilde{\mathsf{P}}_{xy}$ against the azimuth angle $\phi$ so that the amplitude of these curves give us the `worst case' value for the error. The dashed lines show the measured value of $\tilde{\mathsf{P}}_{xy}$ for each Bragg peak. The two phases are shown in different colors: LT is blue and MT is red.}
\end{figure*}

\begin{figure*}
	\centering
	\includegraphics[width=.95\linewidth]{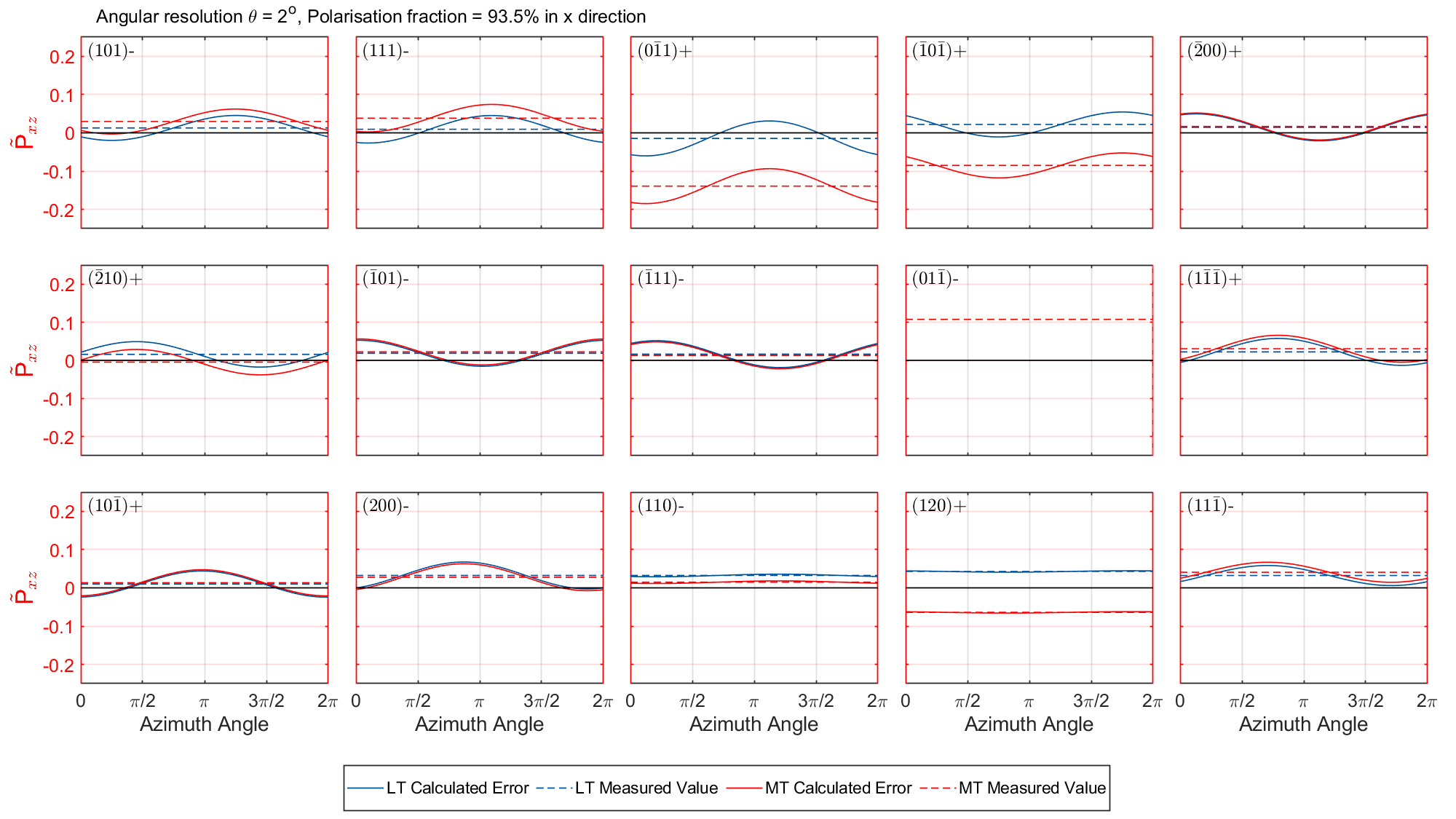}
\caption{\label{fig:PxzError} Figure showing the calculated value of $\tilde{\mathsf{P}}_{xz}$ when an angular resolution of 2$^{\circ}$ is assumed on the incident neutron polarization. The 15 Bragg peaks considered in this study are included an labeled. The solid curves indicate $\tilde{\mathsf{P}}_{xz}$ against the azimuth angle $\phi$ so that the amplitude of these curves give us the `worst case' value for the error. The dashed lines show the measured value of $\tilde{\mathsf{P}}_{xz}$ for each Bragg peak. The two phases are shown in different colors: LT is blue and MT is red.}
\end{figure*}

\begin{figure*}
	\centering
	\includegraphics[width=.95\linewidth]{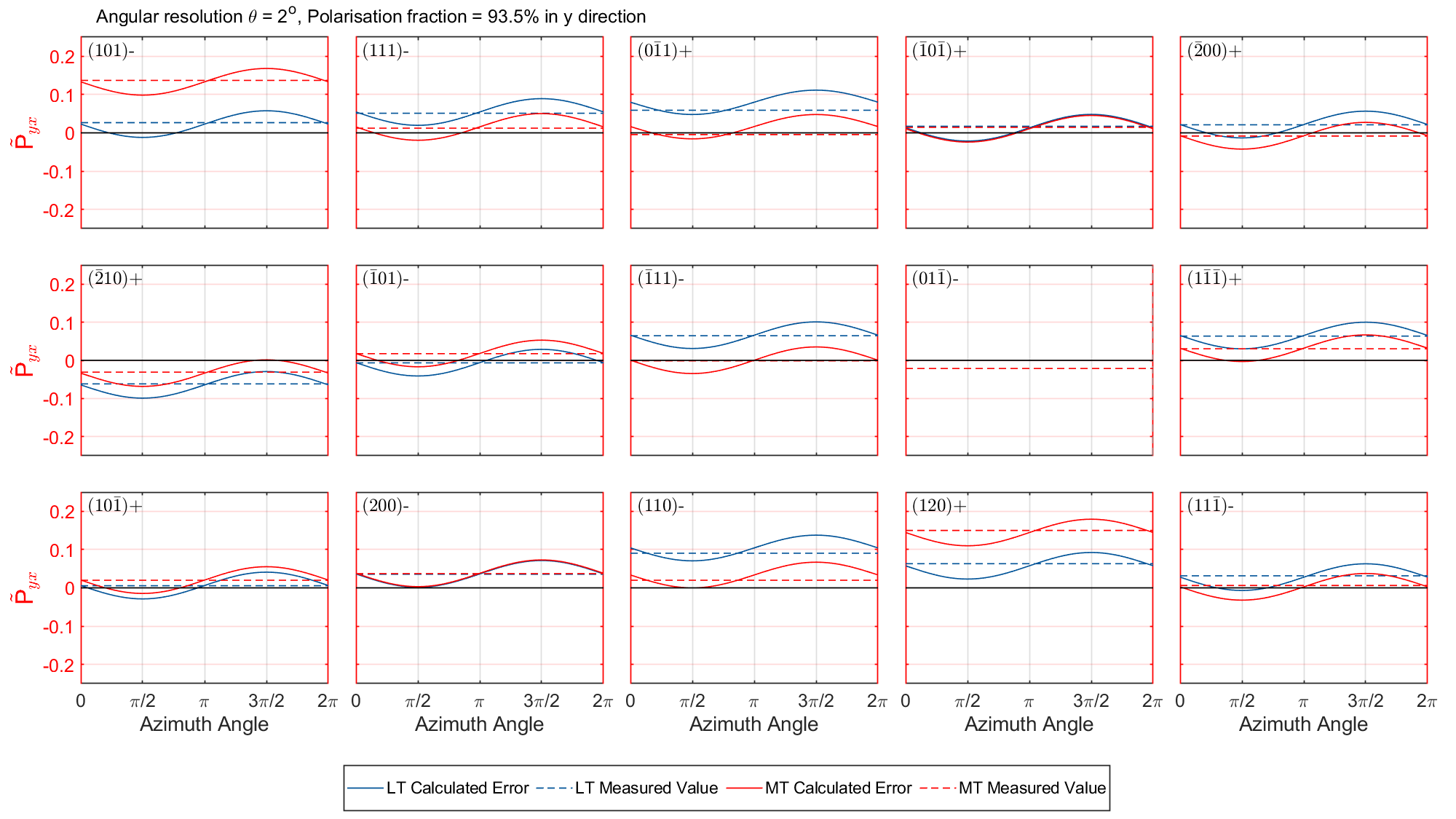}
\caption{\label{fig:PyxError} Figure showing the calculated value of $\tilde{\mathsf{P}}_{yx}$ when an angular resolution of 2$^{\circ}$ is assumed on the incident neutron polarization. The 15 Bragg peaks considered in this study are included an labeled. The solid curves indicate $\tilde{\mathsf{P}}_{yx}$ against the azimuth angle $\phi$ so that the amplitude of these curves give us the `worst case' value for the error. The dashed lines show the measured value of $\tilde{\mathsf{P}}_{yx}$ for each Bragg peak. The two phases are shown in different colors: LT is blue and MT is red.}
\end{figure*}

\begin{figure*}
	\centering
	\includegraphics[width=.95\linewidth]{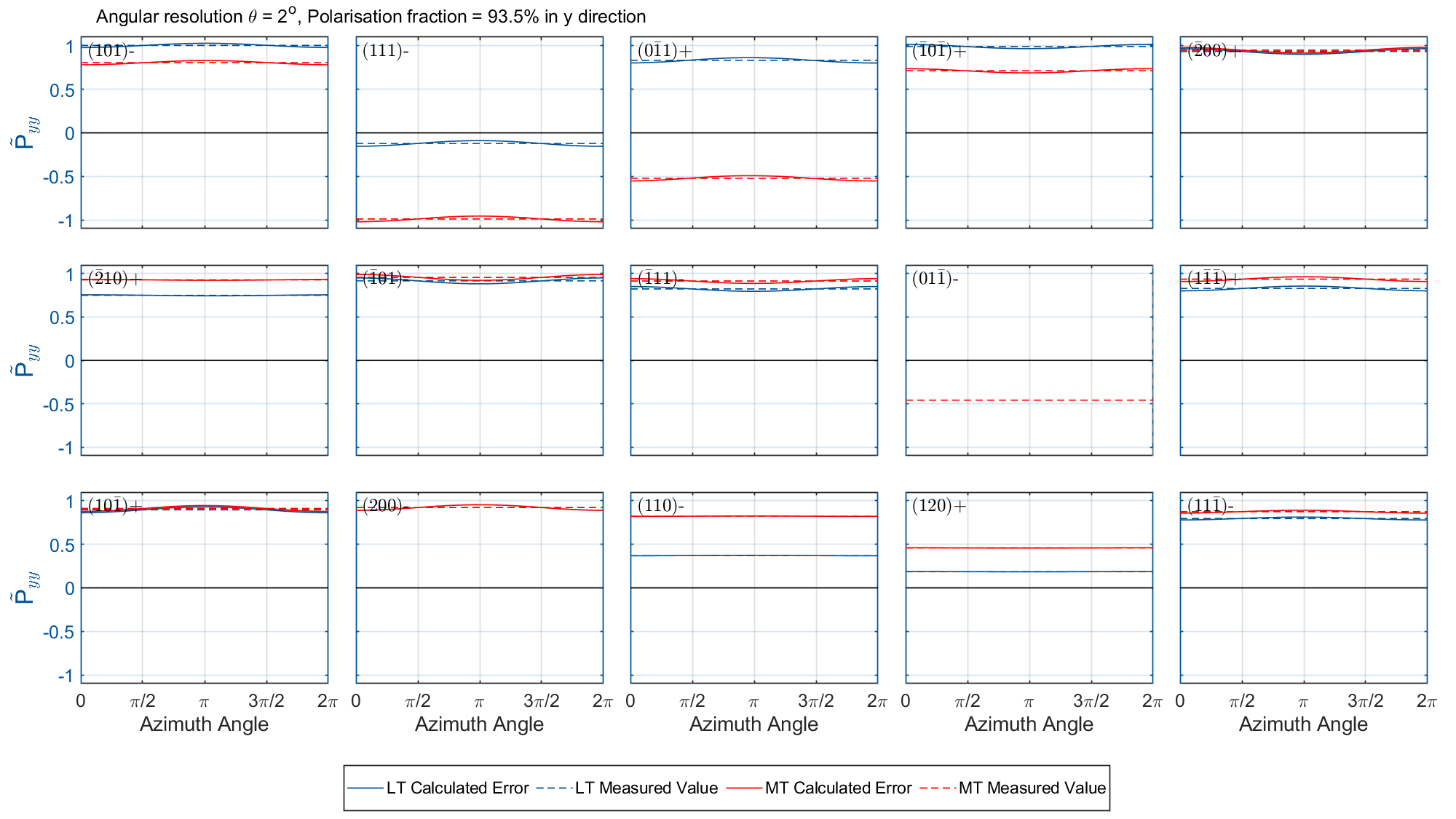}
\caption{\label{fig:PyyError} Figure showing the calculated value of $\tilde{\mathsf{P}}_{yy}$ when an angular resolution of 2$^{\circ}$ is assumed on the incident neutron polarization. The 15 Bragg peaks considered in this study are included an labeled. The solid curves indicate $\tilde{\mathsf{P}}_{yy}$ against the azimuth angle $\phi$ so that the amplitude of these curves give us the `worst case' value for the error. The dashed lines show the measured value of $\tilde{\mathsf{P}}_{yy}$ for each Bragg peak. The two phases are shown in different colors: LT is blue and MT is red.}
\end{figure*}

\begin{figure*}
	\centering
	\includegraphics[width=.95\linewidth]{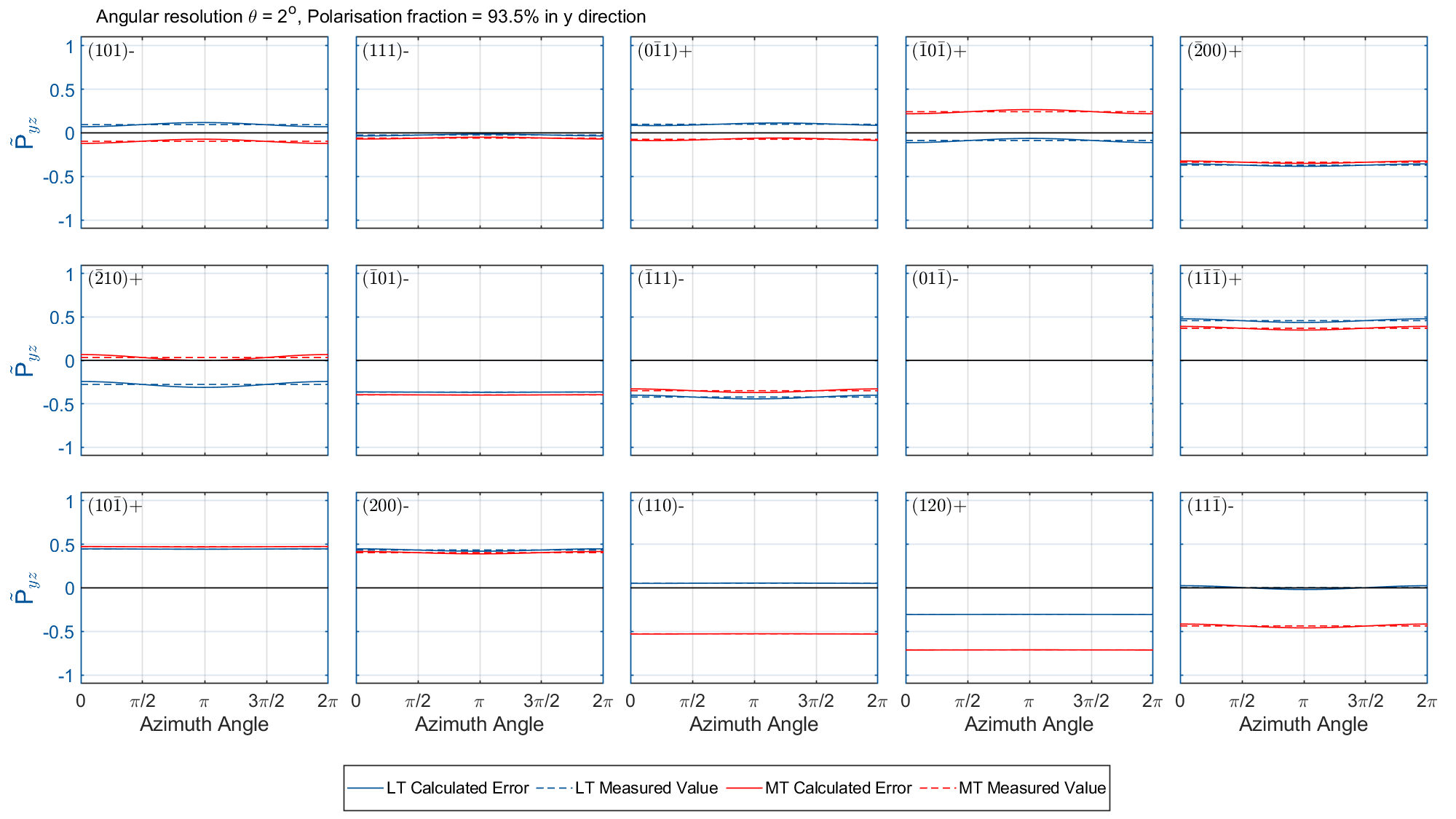}
\caption{\label{fig:PyzError} Figure showing the calculated value of $\tilde{\mathsf{P}}_{yz}$ when an angular resolution of 2$^{\circ}$ is assumed on the incident neutron polarization. The 15 Bragg peaks considered in this study are included an labeled. The solid curves indicate $\tilde{\mathsf{P}}_{yz}$ against the azimuth angle $\phi$ so that the amplitude of these curves give us the `worst case' value for the error. The dashed lines show the measured value of $\tilde{\mathsf{P}}_{yz}$ for each Bragg peak. The two phases are shown in different colors: LT is blue and MT is red.}
\end{figure*}

\begin{figure*}
	\centering
	\includegraphics[width=.95\linewidth]{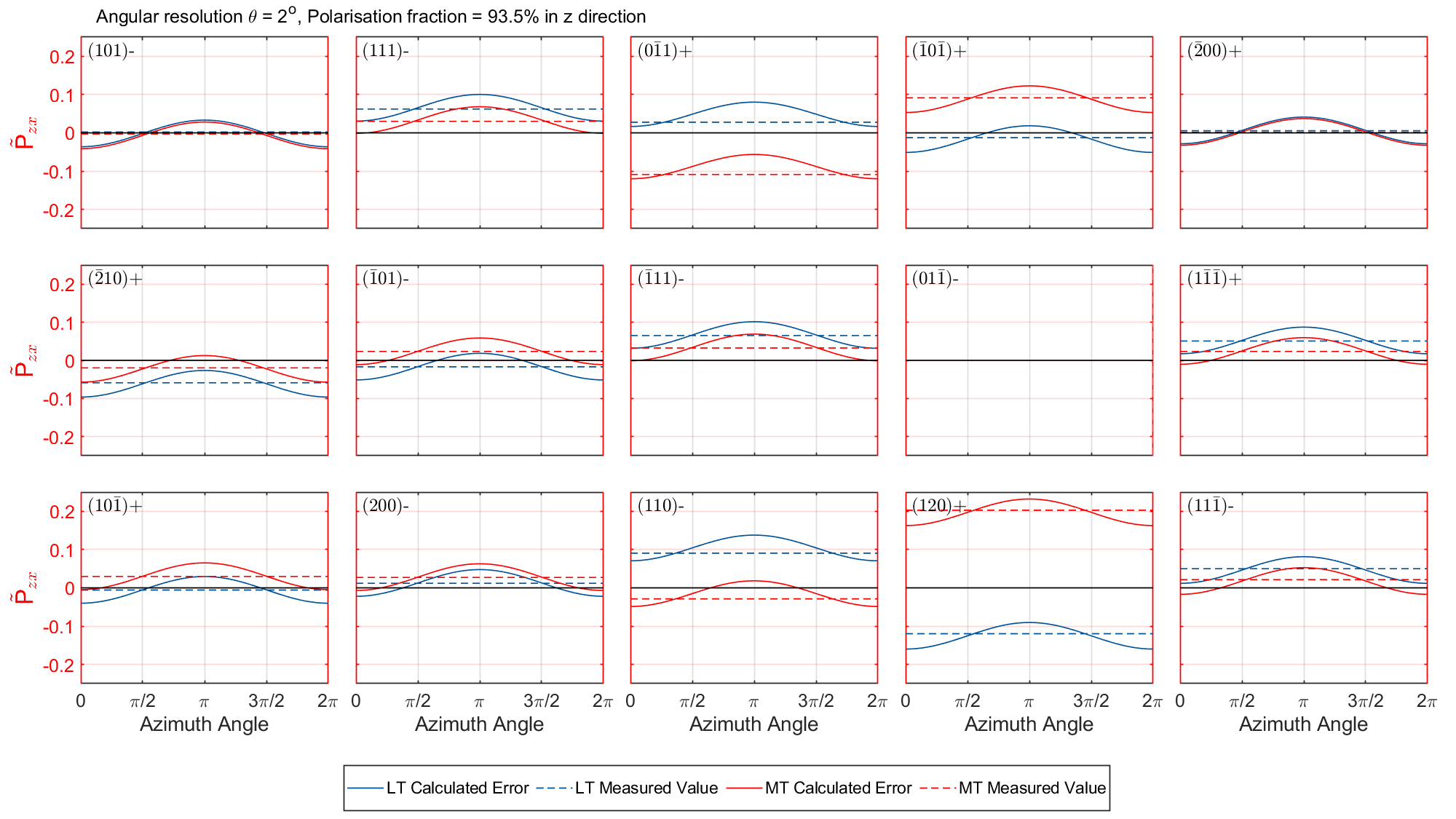}
\caption{\label{fig:PzxError} Figure showing the calculated value of $\tilde{\mathsf{P}}_{zx}$ when an angular resolution of 2$^{\circ}$ is assumed on the incident neutron polarization. The 15 Bragg peaks considered in this study are included an labeled. The solid curves indicate $\tilde{\mathsf{P}}_{zx}$ against the azimuth angle $\phi$ so that the amplitude of these curves give us the `worst case' value for the error. The dashed lines show the measured value of $\tilde{\mathsf{P}}_{zx}$ for each Bragg peak. The two phases are shown in different colors: LT is blue and MT is red.}
\end{figure*}

\begin{figure*}
	\centering
	\includegraphics[width=.95\linewidth]{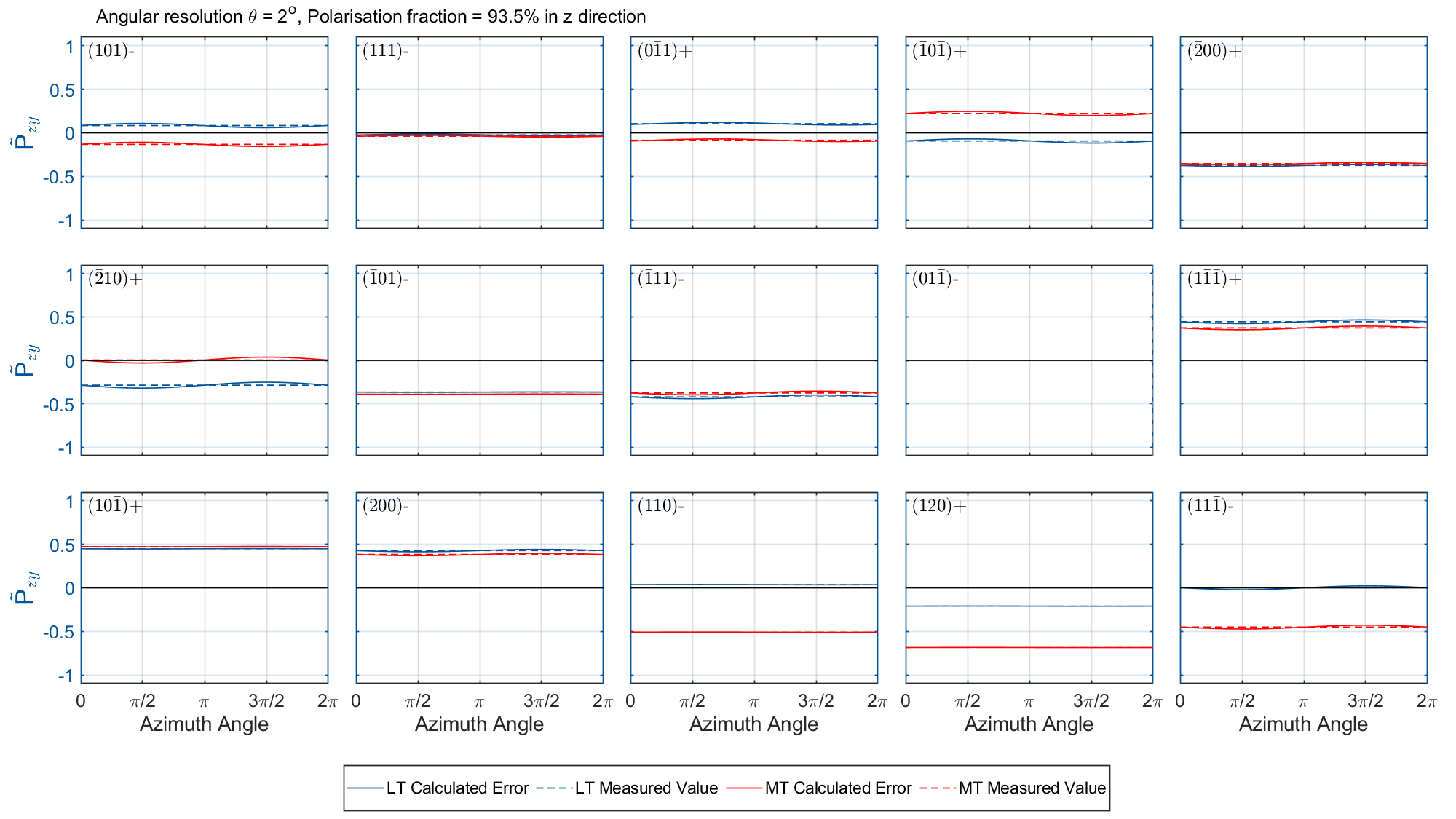}
\caption{\label{fig:PzyError} Figure showing the calculated value of $\tilde{\mathsf{P}}_{zy}$ when an angular resolution of 2$^{\circ}$ is assumed on the incident neutron polarization. The 15 Bragg peaks considered in this study are included an labeled. The solid curves indicate $\tilde{\mathsf{P}}_{zy}$ against the azimuth angle $\phi$ so that the amplitude of these curves give us the `worst case' value for the error. The dashed lines show the measured value of $\tilde{\mathsf{P}}_{zy}$ for each Bragg peak. The two phases are shown in different colors: LT is blue and MT is red.}
\end{figure*}

\begin{figure*}
	\centering
	\includegraphics[width=.95\linewidth]{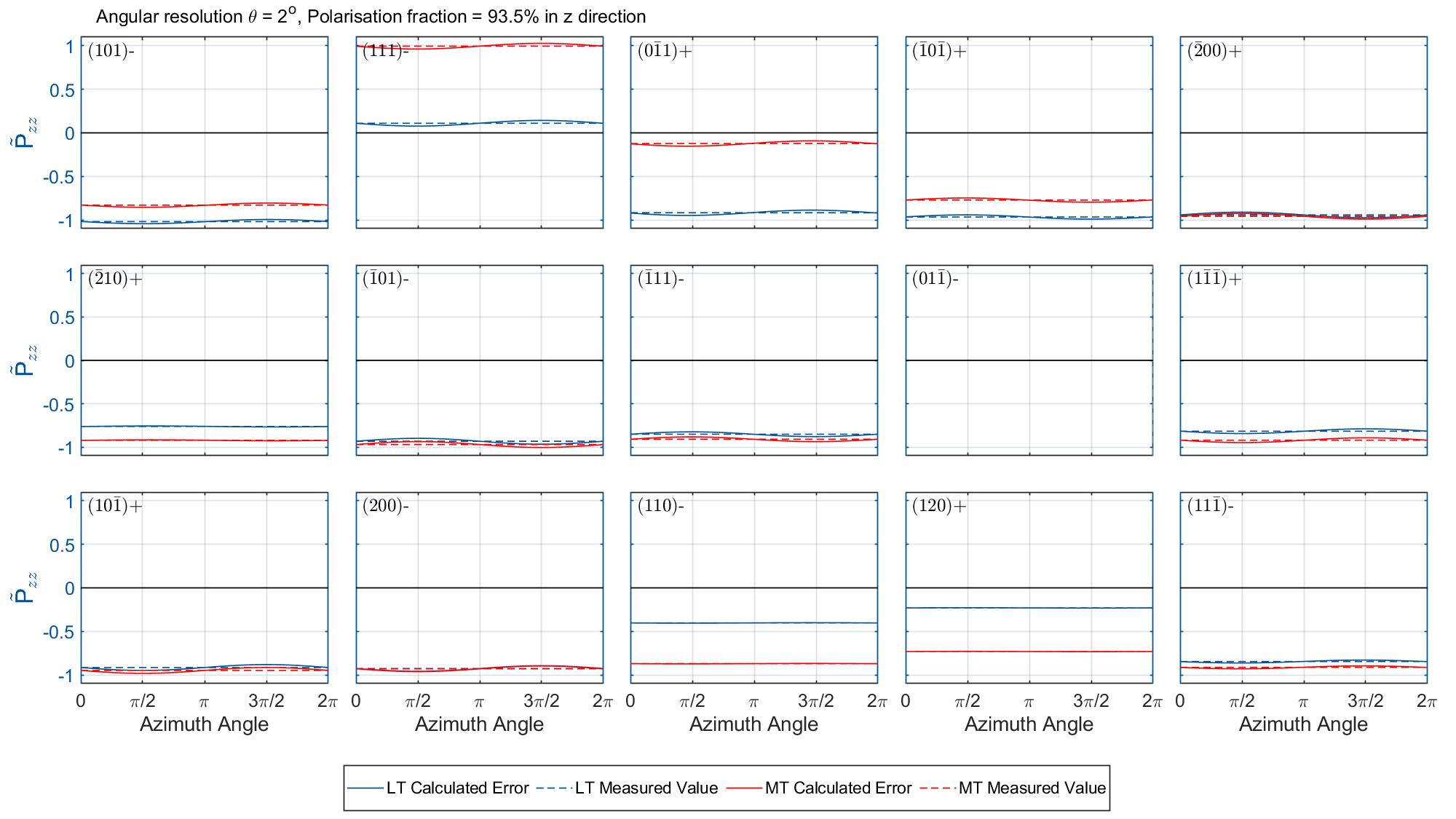}
\caption{\label{fig:PzzError} Figure showing the calculated value of $\tilde{\mathsf{P}}_{zz}$ when an angular resolution of 2$^{\circ}$ is assumed on the incident neutron polarization. The 15 Bragg peaks considered in this study are included an labeled. The solid curves indicate $\tilde{\mathsf{P}}_{zz}$ against the azimuth angle $\phi$ so that the amplitude of these curves give us the `worst case' value for the error. The dashed lines show the measured value of $\tilde{\mathsf{P}}_{zz}$ for each Bragg peak. The two phases are shown in different colors: LT is blue and MT is red.}
\end{figure*}

It is reported\cite{CryopadPaper} that the angular resolution on CRYOPAD is $\approx 2^{\circ}$ and that the initial polarization fraction is $93.5\%$. Setting these values for $\theta$ and $\zeta$ respectively allow us to compute values for Equations \ref{eq:errorBegin} - \ref{eq:errorEnd}. The refined matrix elements were used for this calculation and Fig. \ref{fig:PxxError} - \ref{fig:PzzError} show these errors combined with the statistical ones as a function of azimuth angle $\phi$ for each matrix element. The magnitude of the absolute value of these oscillatory curves give the error. The resultant errors were subsequently included in Fig. \ref{fig:FitsL} and Fig. \ref{fig:FitsM}.


\begin{figure*}[t!]
	\includegraphics[width=.95\linewidth]{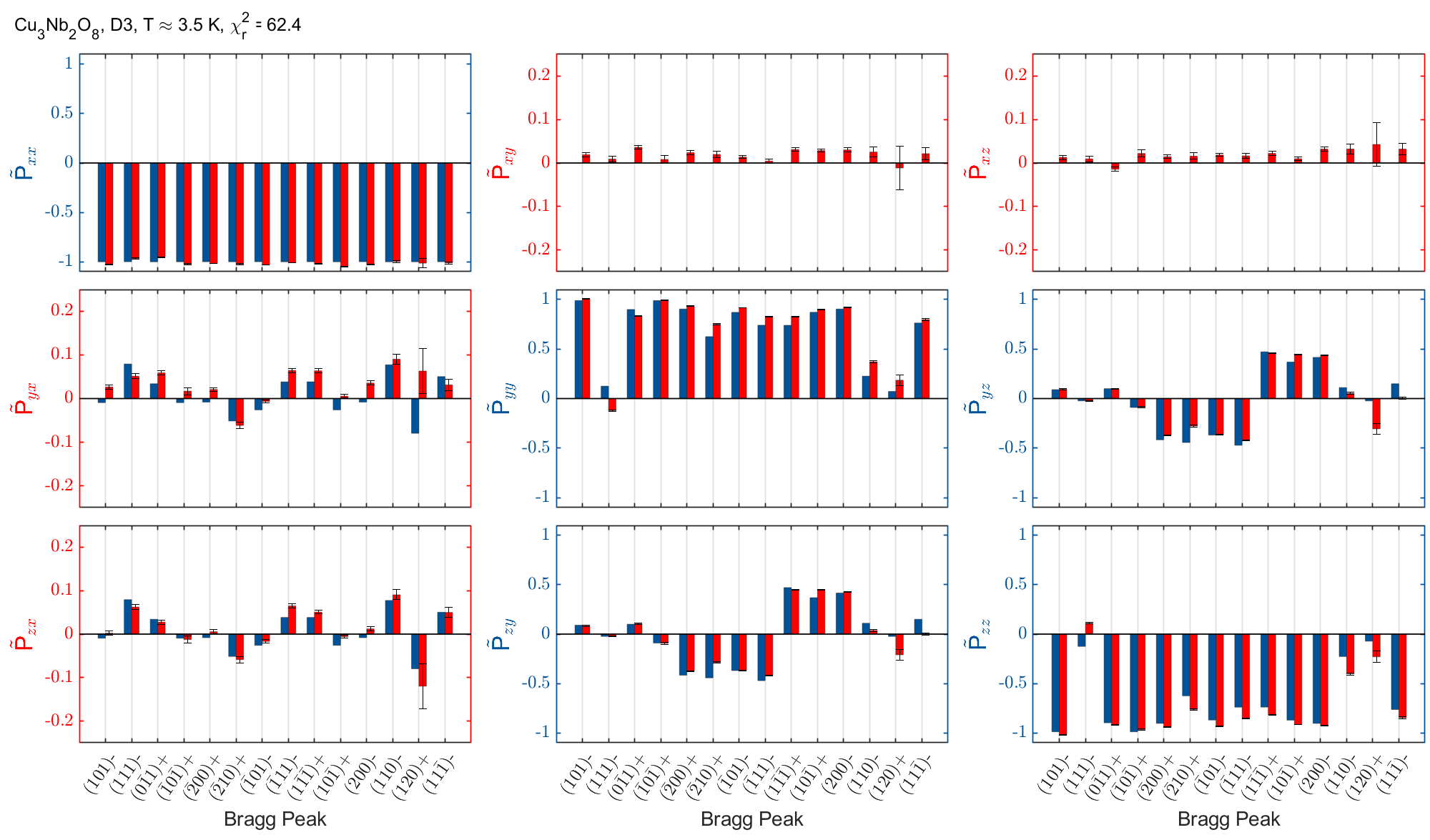}
\caption{\label{fig:FitsLCirc}Figure showing the refinement (in \textsc{Mag2Pol}) of the polarization matrix at $\approx 3.5 K$ -  LT phase. This refinement is the result of constraining the rotational envelope of the Cu$^{2+}$ moments to be circular. This produced a noticeably worse fit than that presented in the main text (section \ref{MagStrucSec}). The bars show the refined matrix elements (left - blue in color) plotted for each Bragg peak against the measured matrix elements (right - red in color). Statistical experimental errors \emph{only} are shown in black. The plotted matrix elements are corrected for detector spin filter efficiency. For clarity, two different $y$ scales are used and are displayed in different colors.}
\end{figure*}

\setcounter{equation}{0}
\renewcommand{\theequation}{B\arabic{equation}}
\section*{Appendix B: Constrained Low Temperature Fit}

In the LT phase a generic helicoid with an elliptical envelope was refined from the SNP data. However, Johnson \textit{et al} reported that their structure had a circular rotation envelope.\cite{Johnson_CNO_Paper} In order to check this, a constraint was added into the refinement process in the LT phase such that the real and imaginary parts of $\vec{M}_{\perp}(\vec{Q})$ were of equal magnitude. Recall that the spin structure is given by

\begin{equation} \label{eqn:Spin}
        S_i(\vec{L})= \mathcal{R}_i \cos(\vec{k} \cdot \vec{L} + \Phi_i) + \mathcal{I}_i \sin(\vec{k} \cdot \vec{L} +  \Phi_i),
\end{equation}

\noindent where $i$ labels the Cu sites and $\vec{L}$ is a real space lattice vector. Coordinates in this section are given with respect to a spherical polar coordinate system $(r,\phi,\theta)$ constructed inside an orthonormal basis $(x^{\prime}y^{\prime}z^{\prime})$ where $x^{\prime} \parallel a$ with $b$ in the $x^{\prime}-y^{\prime}$ plane. We see that if $\mathcal{R}_i$ and $\mathcal{I}_i$ have equal magnitude then the rotation envelope is indeed circular.

However, this produced a worse refinement than that presented in section \ref{MagStrucSec} with $\chi_r^2 = 62.40$  which is must larger than that presented in the main text ($\chi_r^2 = 37.98$). This refinement is shown in Fig. \ref{fig:FitsLCirc} however the errors shown are purely statistical as this fit was not included in the systematic error calculations detailed in Appendix A. However, we can expect that this deviation is due to thermal fluctuations and that this circular limit should be recovered at zero temperature.

\end{document}